\documentclass[preprint,12pt]{elsarticle}

\usepackage{amssymb}
\usepackage{amsmath}

\usepackage{longtable}

\usepackage{booktabs}

\usepackage{lscape}

\usepackage[table]{xcolor}

\usepackage{tabularray}
\usepackage{float}
\usepackage{pgf}
\usepackage{orcidlink}

\usepackage{lineno}

\journal{Arxiv}

\begin{document}

\begin{frontmatter}



\title{Do Resilience Metrics of Water Distribution Systems Really Assess Resilience? A Critical Review}

\author{\texorpdfstring{Michaela Leštáková\corref{cor1}}{Michaela Leštáková} \orcidlink{0000-0002-5998-6754}}
\cortext[cor1]{Corresponding Author}
\ead{michaela.lestakova@fst.tu-darmstadt.de}

\author{Kevin T. Logan \orcidlink{0000-0001-5512-2679}}
\author{Imke-Sophie Rehm \orcidlink{0000-0001-9751-3934}}
\author{Peter F. Pelz \orcidlink{0000-0002-0195-627X}}
\author{John Friesen \orcidlink{0000-0003-2530-1363}}

\affiliation{organization={Technical University of Darmstadt, Chair of Fluid Systems},
            addressline={Otto-Berndt-Str. 2},
            city={Darmstadt},
            postcode={64287},
            state={Hesse},
            country={Germany}}

\begin{abstract}
Having become vital to satisfying basic human needs, water distribution systems (WDSs) are considered critical infrastructure. They are vulnerable to critical events such as extreme weather, natural and man-made disasters, armed conflicts etc. To account for critical events during the design and operation of WDSs, the concept of resilience is frequently mentioned. How resilience of WDSs can be assessed using resilience metrics has been the subject of research of many publications.  The aim of this paper is to inspect the alignment between a general understanding of resilience in WDSs and the metrics used for resilience assessment. A novel framework for categorising resilience metrics for WDSs is presented. A literature review of resilience metrics for WDSs is performed and the results are analysed using the framework designed.
The results show that resilience metrics do not really assess resilience of the systems, but rather only specific functions and properties of systems which can make them resilient.

\end{abstract}



\begin{keyword}
resilience \sep water distribution systems \sep resilience metrics \sep review \sep infrastructure resilience


\end{keyword}

\end{frontmatter}







\section{Introduction}
\label{sec:Introduction}
Access to safe water belongs to the most fundamental human needs \cite{UnitedNationsSustainableDevelopment.1142022}. It plays a pivotal role in the Sustainable Development Goal 6 \cite{UNDESA.July2022}.
In many places on Earth, access to safe water is provided by water distribution systems (WDSs). Having become vital to satisfying basic human needs, WDSs are considered critical infrastructure that is vulnerable to extreme weather events, natural and man-made disasters, armed conflicts etc. Recent examples of this include the disruption of water supply as a consequence of the 2021 floods events in western Europe e.g. in Bad Münstereifel  \cite{Koks.2022}, several cases of direct attacks at pumping stations, pipelines and dams during the Russia–Ukraine armed conflict \cite{Shumilova.2023} as well as broken water pipes as a consequence of the 2023 earthquake in Turkey and Syria \cite{MiddleEastEye.3282023}. The projected increase in frequency of extreme weather events as a result of the progressing climate crisis also affects and will continue to affect WDSs.

To account for critical events in the context of design and operation of WDSs, the concept of resilience is frequently mentioned \cite{Ulusoy.2018}. WDSs are considered technical or socio-technical systems that need to be \textit{resilient} with regard to critical events. However, it is challenging to operationalise the concept of resilience in WDSs and to use it in academic studies \cite{Fekete.2020}. The reasons for this are mainly twofold: (i) no scientific consensus regarding the definition of resilience exists; and (ii) no scientific consensus regarding measuring WDS resilience exists.

These two challenges have been addressed in numerous scientific publications. In particular, numerous metrics have been proposed for resilience assessment of WDSs. The aim of the presented publication is to inspect the alignment between a general understanding of resilience in WDSs and the metrics used for resilience assessment. Specifically, the following research questions are addressed:

\begin{itemize}
    \item How do existing WDS resilience metrics assess resilience?
    \item To what extent do the existing metrics assess resilience with regard to the functions and properties of resilient systems?
    \item How general are the existing resilience metrics with regard to different critical events?
\end{itemize}

To answer these questions, the rest of the paper is structured as follows. First, an overview of existing review papers about resilience metrics in WDSs is provided in Section \ref{sec:Reviews}. In Section \ref{sec:Resilience}, the understanding of resilience within the scope of this study is
presented, placed in the overall resilience discourse and its implications
for the WDSs are illuminated. The novel framework for classifying resilience metrics is described in Section \ref{sec:Framework}. Section \ref{sec:Search} documents the literature search protocol. The results of the critical review and the discussion are presented in Section \ref{sec:Results} and \ref{sec:Discussion}, respectively.

\section{State of the Art}
\label{sec:Reviews}
Several review studies aimed to categorise resilience metrics for water distribution systems in the past: Liu and Song.~\cite{Liu.2020}, Shuang at al.~\cite{Shuang.2019} and Shin et al.~\cite{Shin.2018}, Gunawan, Schultmann and Zarghami \cite{Gunawan.2017}, Gay and Sinha \cite{Gay.2013}, and Mohebbi et al. \cite{Mohebbi.2020}.
Each of the studies used their own unique framework for categorising metrics and the understanding of resilience also varied. In the following, the key structure of each of the frameworks is presented.

Shin et al.~\cite{Shin.2018} broaden the horizon considering not only WDSs but also water resource systems (WRSs).
While stating that resilience definitions in the domain of water infrastructures lack clarity, they determine four key capabilities of resilient systems - withstanding capability, absorptive capability, restorative capability and adaptive capability.
These capabilities are considered as customer needs in a functional design process and can thus be understood as system functions.
Shin et al. categorise resilience metrics according to two separate dichotomies: probabilistic vs. deterministic and dynamic vs. static.
Unlike deterministic measures, the probabilistic measures "consider the stochasticity of system functions (or disturbances) and the probability-based formulation of the measures"\cite{Shin.2018}.
Dynamic approaches "consider time-dependent functions of a system" while time-independent approaches do not.

Focusing solely on WDNs, Shuang et al.~\cite{Shuang.2019} define WDN resilience as the "ability to absorb local failures, to quickly recover and maintain the essential service functions, and to adapt to long-term changes in the environment and uncertainty disturbances" \cite{Shuang.2019}.
From this definition, they abstract three capabilities of a resilient WDN: absorptive, restorative and adaptive, omitting the withstanding capability of Shin et al. \cite{Shin.2018}.
Analysing existing publications related to resilience assessment of WDNs, the authors identify four clusters of approaches for quantitative resilience metrics: surrogate measures, simulation methods, network theory approaches and fault detection and isolation approaches \cite{Shuang.2019}.
For each of the approaches, an overview of metrics, research progresses and limitations is provided.
The clusters are, however, qualitatively different: while the first three focus on the methods behind the metrics, the last one covers an application area.

Liu and Song reviewed the body of research carried out on WDSs and five different types of urban networks (drainage distribution, gas distribution, transportation, electricity distribution and communication) \cite{Liu.2020}.
For WDSs, Liu and Song identify  two types of metrics similar to those of Shuang et al. \cite{Shuang.2019}: surrogate-based evaluation metrics and recovery-based simulation metrics.
According to the authors, the definition of resilience also changes based on which of the two types of metrics is used: in the first case, resilience "is considered a surrogate measure of [...] reliability, robustness, reserve capacity, and sustainability"\cite{Liu.2020} and is thus "static"\cite{Liu.2020}.
In the second case, the resilience definition "[includes] adaptability, absorbability, and recovery capacity"\cite{Liu.2020}, and is a "reflection of dynamic system performance before and after hazards"\cite{Liu.2020}.
However, the authors do not provide any sources for these definitions, neither do they explain their understanding of the terms "static" and "dynamic".
They are also unclear on whether these terms relate to the concept of resilience, the resilience metric or the technical system itself.

Gunawan, Schultmann and Zarghami consider resilience itself to be one of the "indicators of system performance", along with reliability,  redundancy and robustness.
They list 14 metrics, assigning each to one of the four "indicators" and dividing them into structural and functional metrics, where structural metrics can be understood as analogous to static metrics of Shin et al. and functional metrics as analogous to dynamic metrics \cite{Shin.2018}.
Resilience is mentioned only in connection with the dynamic metrics.
Although this analysis provides an initial overview of different metrics, it does not systematically compare the individual indicators and elaborate in detail how they differ or what function of resilience they are related to.

Gay and Sinha performed a literature review of civil infrastructure system's resilience \cite{Gay.2013}. They offer an interdisciplinary perspective, distinguishing between engineering, ecologic, economic and societal resilience. However, they lack a resilience definition for the case of engineering resilience.
They argue that while resilience in general cannot be measured, a system's capability for resilience can be assessed by concepts from graph theory. This resilience assessment should consider the previously stated four aspects of resilience during operation, design and analysis of infrastructures.

Mohebbi et al. \cite{Mohebbi.2021} evaluate resilience and its quantification in water, cyber, and transportation infrastructures, as well as their interdependencies. They distinguish between network-based, performance-based and technology-based metrics. They provide comprehensive lists  for the different metrics (7 network-based, 6, performance-based and 5 technology-based). Nevertheless, similar to the reviews before, the authors do not go into detail about their understanding of resilience and do not describe which aspects of resilience each metric describes.

A major methodological problem in the review studies mentioned above is that little to no effort is made to link the resilience metrics to the definition or understanding of resilience.
"Functions" or "capabilities" of resilient systems are mentioned, but not thoroughly reflected by the categorisation or analysis of the resilience metrics themselves.
Other concepts such as redundancy, reliability and robustness are mentioned, but their relation to resilience differs in each paper and in some cases \cite{Liu.2020, Gunawan.2017} it is unclear whether they are the property of resilience or of a resilient system.
Hence, more work is needed to improve the connection between the interpretation of resilience, other related concepts commonly mentioned in its context and the metrics used to measure it.
The presented paper proposes a framework to address this challenge.

\section{Resilience and Water Distribution Systems}
\label{sec:Resilience}
In this section, the understanding of resilience underlying this study is presented and placed in the overall resilience discourse.
Certain functions and properties of resilient technical systems are introduced and the implications of this understanding within the studied domain of WDSs are illuminated.
\subsection{Resilience of Technical Systems}
\label{subsec:overview}
While earlier mentions of the term resilience can be found, the first usage considered relevant for this work is by C. S. Holling in 1973~\cite{Holling.1973}.
Holling describes resilience as a measure of the ability of ecosystems "to absorb changes of state variables, driving variables, and parameters, and still persist."
In a later work, Holling distinguishes between engineering resilience and ecological resilience~\cite{Holling.1996}.
The former focuses on efficiency, constancy, and predictability and aims for resistance of a (ecological) system to perturbation and return to an equilibrium steady state.
The latter, in contrast, allows for multiple steady states to exist and considers the magnitude of disturbances that cause regimes changes in a system from one state to another.

Since then, the term resilience has been widely adopted and discussed in multiple scientific fields, as indicated by the scope of contributions to the Handbook of International Resilience~\cite{Chandler.2016}.
According to Elsner et al. the concept of resilience owes its popularity in parts to a conjuncture of ecology, awareness of the dynamic nature of systems and the unavoidability of failures as well as a certain fatalism towards a loss of control~\cite{Elsner.2018}.
In consequence, resilience has come under increased critical scrutiny.
One point of criticism is the vagueness and ambiguity of its meaning~\cite{Canizares.2021} or even its haphazard usage~\cite{Elsner.2018}.
This has not only put into question its usefulness for scientific study but raised the concern that as a normative term it transports a hidden agenda, as it does not capture aspects of political and economic power or interests, but instead is in line with neoliberal ideology~\cite{Elsner.2018}.
More explicitly, it has been argued that calling for resilience is a strategy for the shifting responsibility of coping with critical events from large social institutions to individuals and that it can serve as an excuse for inaction with regard to mitigating the consequences of critical events or developments~\cite{Canizares.2021}.
The question whether the term is normative is not fully resolved, however, as the resilience of constellations or systems can be both desirable and undesirable~\cite{Canizares.2021}.
This is also reflected in the metaphors used for describing resilience, in that it allows systems to "bounce back" or "bounce forward".
Here, the former implies that a disrupted system returns to a prior, desirable state, reminiscent of Holling's concept of engineering resilience and the latter that a disruption of the system leads to transformation and a new state of the system reflecting Holling's concept of ecological resilience.

In spite of the critique, it has been acknowledged that the term is useful when studying complex, transient, adaptive systems~\cite{Elsner.2018}.
Accordingly, from Holling's concept of engineering resilience a paradigm of resilience engineering has developed for engineers concerned with complex systems~\cite{Woods.2017, Hollnagel.2011, Hollnagel.2016}.
Within this domain, the definition of resilience for the purpose of engineering of complex systems was gradually and systematically developed in order to include reactions to mishaps or continuous stress, to highlight the uncertainty of these events, and finally to incorporate aspects of the ecological engineering concept by focusing on adaptation to changed conditions~\cite{Hollnagel.2016}.

Inspired by Hollnagel, Pelz et al. drew on resilience as a strategy for coping with uncertainty when designing load-bearing systems in mechanical engineering~\cite{Pelz.2021b}.
Maintaining that systems ultimately serve to fulfil functions, they differentiate between three types of uncertainty these systems face: stochastic uncertainty, incertitude and ignorance.
They further propose three design strategies for coping with uncertainty: (i) robustness, (ii) flexibility, and (iii) resilience.
Here, robust systems are able to fulfil their designed functionality not only at the design point but within a given interval of operating conditions around the design point, whereas flexible systems can adapt to fulfil a given set of predetermined functionalities depending on the operating conditions.
Both strategies are used for coping with incertitude.
Resilience, in contrast, is a strategy for coping with ignorance, as it allows for systems to evolve their function beyond the predefined design point as an adaptation to changed conditions, while still fulfilling at least the function of its initial design.
Accordingly, the authors give the following definition:

\begin{quote}
   A resilient technical system guarantees a predetermined minimum of functional performance even in the event of disturbances and failures of system components, and a subsequent possibility of recovering.~\cite[p.~411]{Pelz.2021b}
\end{quote}

In this conceptualisation, resilient systems are a strict subset of flexible systems, which in turn are a strict subset of robust systems.

Within the scope of the presented work, resilience is understood as the property of technical systems according to the definition given above.
However, the more general term \textit{critical events} is used instead of disturbances or failures to describe any events that require the system to operate outside the designed operating conditions.
In the following subsection, this broad definition is further detailed.

\subsection{Functions and Properties of Resilient Technical Systems}
\label{subsec:functionsandproperties}


Hollnagel speaks of four functions that make resilient performance possible~\cite{ Hollnagel.2016, Hollnagel.2011}. These functions of resilient systems are~\cite{Hollnagel.2016}:

\begin{itemize}
    \item \textbf{monitoring} (knowing relevant internal and external critical parameters; supervising their values during operation)
    \item \textbf{reacting} (being able to respond to critical events by adjusting the current mode of functioning)
    \item \textbf{learning} (understanding what happened during a critical event and incorporate the knowledge during future critical events)
    \item \textbf{anticipating} (knowing the expected system's behaviour when faced with critical events and being able to anticipate future developments such as changing operating conditions or new critical events)
\end{itemize}

\begin{figure}
    \centering
    \includegraphics[width =12cm, trim={4cm 4cm 4cm 4cm},]{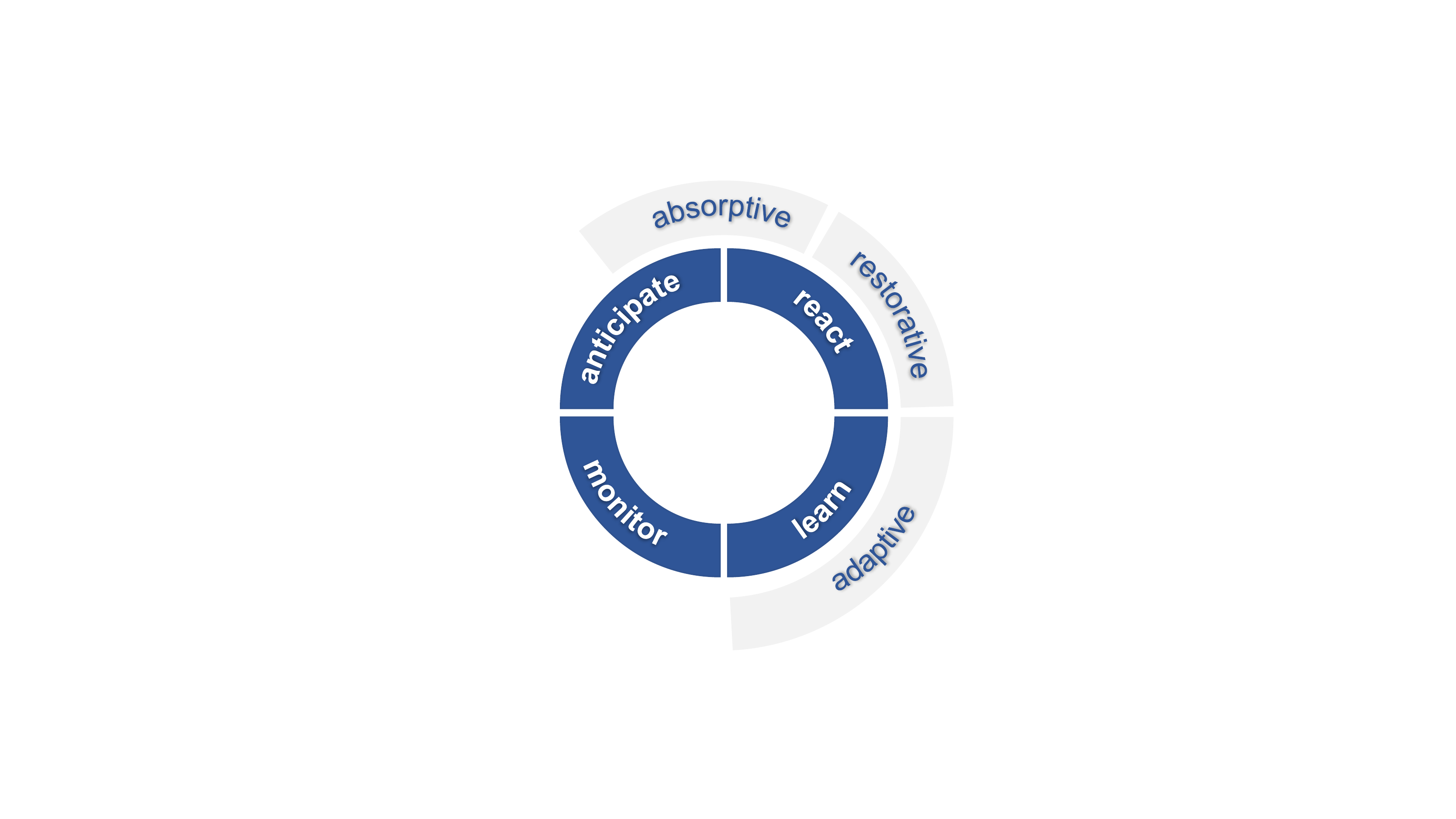}
    \caption{Mapping of capabilities of resilient systems to their functions. While the absorptive capability can be mapped to both ancitipation and reaction, the restorative and adaptive correspond to reaction and learning, respectively.}
    \label{fig:functions_capabilities}
\end{figure}

Resilient systems are often described as those having the  \textit{adaptive}, \textit{absorptive} and \textit{restorative} capability \cite{Liu.2020, Shin.2018, Shuang.2019, Hosseini.2016, Ulusoy.2018}.
These capabilities can be mapped to the functions of resilient systems as shown in Figure \ref{fig:functions_capabilities}.

Besides the functions of resilient systems, several other properties result from the definition of resilience in Section~\ref{subsec:overview}.
Resilient systems are defined as having a predetermined (meaning required, acceptable) minimum of functional performance. This is e.g. the minimum of functional performance for emergency operation during or after a critical event. In resilient systems, the intrinsic minimal functional performance lies above the predetermined functional performance.
It can be considered a baseline, hence it is referred to as \textit{baseline functionality} in the context of this paper.
Another important property of resilient systems is the possibility of \textit{recovery}, reflected in the definition of resilience: despite the functional performance of the system being compromised, it should be possible for the system to return to a state in which a satisfactory functional performance can be guaranteed. Recovery is present in many other definitions of resilience as well \cite{Ulusoy.2018}.
A further property of resilient systems is \textit{redundancy} \cite{Ulusoy.2018}: by equipping the system with additional capacity, e.g. with duplicate components, a sufficient functional performance can be secured even in case of a critical event.
Redundancy is generally considered tightly coupled with the concept of, but not sufficient for resilience. As such, it accompanies baseline functionality and the possibility of recovery.

While the concept of resilience and the functions and properties of a resilient technical system have so far been presented in abstract terms of systems in general, in the following subsection, they are concretised for WDSs.

\subsection{Resilience Applied to Water Distribution Systems}



WDSs are large technical systems (LTS) with a socio-political dimension~\cite{Forster.2015, Moss.2020}.
As infrastructure systems, they lie within the engineering domain as engineering knowledge is required for their design and operation \cite{Hosseini.2016}.
In the context of this work, the focus is on the technical character of WDSs.
Of the overall water supply system of a city, WDSs are defined as the part that transports water from the outlet of the source or treatment plant to the point where the consumer's installation connects~\cite{EuropeanCommitteeforStandardization.April2022}.
WDSs consist of a network of pipes of various carrying capacity that stretches out covering the supply area as well as service reservoirs, pumping stations, valves, joints and fittings and further minor components~\cite{EuropeanCommitteeforStandardization.April2022}.

Considering the adopted definition of resilience in the context of WDSs, the concepts minimum of functional performance, failures of system components, disturbances, and possibility of recovery are to be clarified.

Functional performance of WDSs is determined by threshold values at the point of connection to the consumer's installation for the following quantities: service pressure, flow rate, continuity of supply, water quality (i.e. maximum threshold values for substances in the water)~\cite{EuropeanCommitteeforStandardization.April2022}.
Further criteria for functional performance include sustainable use of energy, minimising water loss, longevity of installations, minimising noise, and minimising risks to neighbouring buildings and the environment, and providing service in emergencies~\cite{DINNormenausschussWasserwesen.February2021}.

The minimum of functional performance can be defined by a further set of threshold values for the quantities enumerated above, according to national regulations.
As an example for volume of water, the Federal Office of Civil Protection and Disaster Assistance (BBK) gives an estimation of $50$ litres per day and capita to be provided by operators of WDS, even during critical events~\cite{BundesamtfurBevolkerungsschutzundKatastrophenhilfe.22.02.2022}.
This figure corresponds to the level 6B water restrictions enforced by the City of Cape Town during the drought in 2018~\cite{WesternCapeGovernment.20.04.2023}.
\footnote{The amount of water provided in emergencies may be lowered to the minimum of water required by humans as defined by humanitarian NGOs or governmental organisations tasked with civil protection, which is $15$ litres per day and capita~\cite{McCann.2018, BundesamtfurBevolkerungsschutzundKatastrophenhilfe.22.02.2022}. This is generally not distributed through the technical WDS but by different means, which is why this threshold value is unsuitable as a minium of functional performance for the resilience assessment of WDSs.}
The threshold value in this case is a requirement defined for the operation of the WDS as a baseline functionality and is not equivalent to a predetermined minimum of functional performance as a characteristic of the WDS.
Since WDSs are rarely designed from scratch but rather develop over generations, a predetermined minimum of functional performance cannot be implemented as a system characteristic and determining this characteristic is not trivial.
Thus, defining threshold values for an acceptable minimum of functional performance for all operating conditions (i.e. a baseline functionality) is usually more relevant than determining the actual system  characteristic "minimum of functional performance" of a WDS.

Failures of system components in WDSs include but are not limited to pipe breaks, leakages in pipes, joints, service reservoirs or fittings, faults in pumping stations and pump outages, and broken valves.
Disturbances are considered to be changes to the operating condition deviating from that for which the WDS was designed, without physical damages to components.
This includes unexpected changes to consumer demand and demand patterns, changes to the available supply of water from the sources and treatment plants as well as contamination of the water sources, back flow, and stagnation.
The union of failures of system components and disturbances is termed critical events in the context of this work.

Concerning the possibility of recovery subsequent to critical events, this can generally be understood as the WDS returning to the service levels determined by the functional performance after a period in which only at least the minimum of functional performance was fulfilled.
Depending on the nature of the critical events, recovery is either achieved from within the WDS through the actuators (pumping stations, valves) or by human intervention (repair of pipes and other broken components, restoring supply through source or water treatment plant, and others).
It is important to recognise that in the latter case, the system boundary is extended to include not only the WDS as described above but also the human agents required to operate it as well as spare materials.

The four resilience functions defined by Hollnagel referenced in the preceding section are proposed to be understood in the context of WDSs as follows:
\begin{itemize}
    \item \textbf{monitoring} using sensors to measure quantities for operation, e.g. service pressure and volume flow, as well as relevant external quantities, e.g. groundwater levels, precipitation, population dynamics
    \item \textbf{reacting} mitigating the effects of critical events on functional performance after detecting them and returning the WDS to fulfilling service levels, e.g. using actuators in the WDS such as valves and pumps (when not including human agents in the system boundary) or deploying repair crews to restore failed components (when including human agents in the system boundary)
    \item \textbf{learning} gathering operation data and information and analysing them to improve future operation. This can entail using the data to adapt models for control units of pumps or improving protocols for detecting critical events through monitoring, e.g. by improving data analysis methods of the sensor data
    \item \textbf{anticipating} providing for likely critical events in WDSs, e.g. leakages, pipe breaks, pump outages, demand or supply variations, and others, as well as considering long term developments, e.g. demand level increase or decrease through migration into or out of the supply area and changes in supply due to dropping groundwater levels or droughts
\end{itemize}

Considering the properties of resilient technical systems given in the previous section, the following concretisations can be made in the context of WDSs.
\textit{Baseline functionality} is a set of threshold values of the functional performance of the WDS (service pressure, flow rate, etc.) that serves as a reference for the acceptable minimum of functional performance. The deviation from this can be measured.
\textit{Recovery} is reflected in the resilience definition and is closely linked to the resilience function \textbf{react}.
If the threshold values of functional performance cannot be maintained due to critical events, the react function of the WDS needs to be fulfilled in order to return the operating point to a state where the threshold values are again met.
\textit{Redundancy} in WDSs is related to the resilience function \textbf{anticipate}.
Redundancy in WDSs is achieved by, e.g., ensuring multiple supply paths for consumers in a network, using multiple sources, including surplus pumps in pumping stations as well as securing extra capacity both in terms of available volume of water and transportation capacity in the pipe network.\\
Having illustrated how the resilience definition, resilience functions and related resilience properties can be applied to WDSs as an instance of resilient technical systems, the next step is to construct a framework within which metrics for measuring resilience can be classified.

\section{Framework for Classifying Resilience Metrics}
\label{sec:Framework}
In this section, the categories used within the presented work for classifying resilience metrics are presented in detail, constituting the framework used for analysing resilience metrics.

In the past years, a plethora of metrics have been designed for the purpose of assessing resilience of water distribution systems. To inspect \textit{how} the metrics approach the assessment of resilience, the current section presents a framework to classify them according to:
\begin{itemize}
    \item system functions addressed (cf. Section~\ref{subsec:functionsandproperties})
    \item system properties addressed (cf. Section~\ref{subsec:functionsandproperties})
    \item dependence on time (cf. Section~\ref{subsec:time_dep})
    \item mathematical characteristics (cf. Section~\ref{subsec:mat_char})
    \item quantification type (cf. Section~\ref{subsec:quant_type})
    \item scope of the metric (cf. Section~\ref{subsec:scope})
\end{itemize}

An important distinction is that the first two categories refer to what characteristics of the system are addressed, while the rest of the categories are characteristics of the metrics themselves. Hence, the metric \textit{assesses} a function or a property that a system \textit{has}, but the metric \textit{is} time-dependent or \textit{has} certain mathematical characteristics.

As the system functions and properties have already been presented in detail in Section~\ref{subsec:functionsandproperties}, this section will focus on the remaining categories of the framework.

\begin{figure}[!ht]
    \centering
    \includegraphics[width=12cm, trim={4cm 0cm 4cm 0cm},clip]{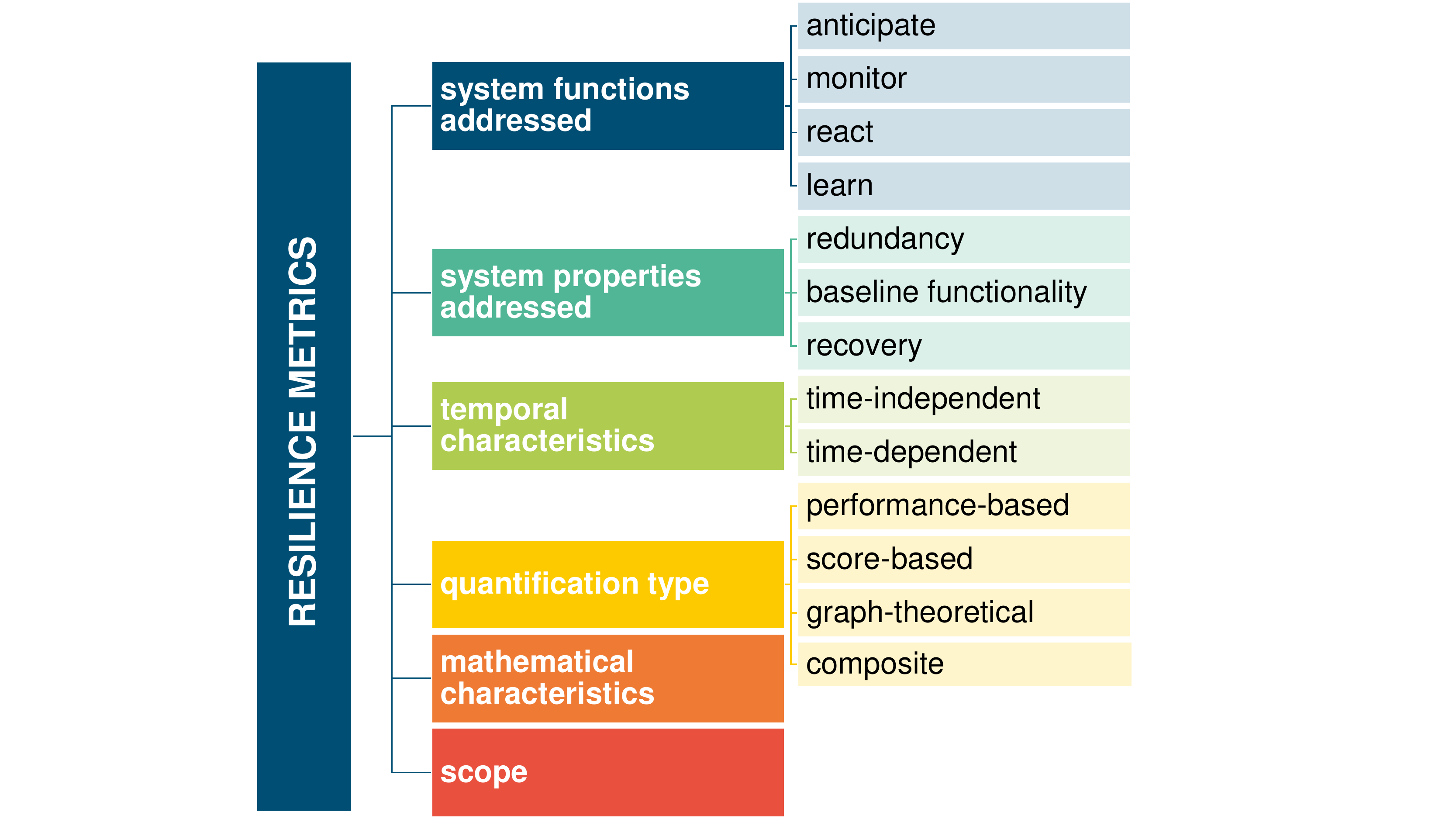}
    \caption{Visualisation of the classification of metrics}
    \label{fig:classification}
\end{figure}

\subsection{Metrics According to Their Dependence on Time}\label{subsec:time_dep}
Resilience metrics can be differentiated based on whether they do or do not consider development of the functional performance of the WDS in time \cite{Liu.2020, Shin.2018, Hosseini.2016}. In this framework, the terms \textit{time-independent} and \textit{time-dependent} are used instead of static and dynamic \cite{Shin.2018} (as the metric itself is not static or dynamic) or structural and functional \cite{Gunawan.2017} (as metrics assessing structural or functional characteristics of a system can still either depend or not depend on time).

\textit{Time-independent} resilience metrics aim to assess resilience without considering the development of the selected quantity in time and tend to focus on topology  of the system and characteristics of its components.

\textit{Time-dependent} resilience metrics account for the development of the functional performance or another selected quantity in time.

It is important to distinguish between time dependence as the property of resilience versus the property of the resilience metrics. In the understanding of the authors of this paper, it is the metrics that can be either time-dependent or time-independent, not resilience itself. Some authors speak of "static resilience" \cite{Sweya.2021}, which would suggest its time independence. However, the authors of this paper are of the opinion that in such case, it would be more appropriate to discuss whether resilience is \textit{time-invariant}. This discussion is, however, beyond the scope of the present work.

\subsection{Metrics According to Their Mathematical Characteristics} \label{subsec:mat_char}
Resilience metrics are defined on various intervals.
Unlike open intervals, closed intervals with an optimal value suggest that a WDS can achieve absolute resilience.
However, no scientific consensus exists about whether this is possible.
The main reason for this is that the resilience scholarship tends to think of resilient systems with regard to any (reasonable) critical events, not to a specific set of them \cite{Mentges.2023}, and that it is impossible to account for all of these in the resilience analysis.
Moreover, it is also disputed whether resilience is a continuous or a Boolean property: whether a system can be \textit{only a little resilient} or whether it either is resilient or is not.

Resilience metrics are developed with the goal of being able to compare various configurations of a single system or separate systems with one another.
Resilience metrics normalised to a closed interval (such as $[0,1]$) suggest that comparability within the system as well as between various systems is possible.
Non-normalised metrics make comparison between separate systems more difficult.



\subsection{Metrics According to Their Quantification Type} \label{subsec:quant_type}
Resilience metrics use different types of quantification. Cassottana et al. differentiate between \textit{graph-theoretic} and \textit{performance-based} resilience metrics \cite{Cassottana.2021}. \textit{Graph-theoretic} resilience metrics are based on measures developed in graph theory \cite{Cassottana.2021}, such as betweenness centrality or shortest paths. As WDSs can be modelled as mathematical graphs, these metrics are a suitable tool for their resilience assessment. Graph-based metrics often aim to express resilience in terms of values of each node or link.
\textit{Performance-based} resilience metrics assess resilience as based on a system output characterising the performance of the system \cite{Cassottana.2021}. For example, they express the ratio of functional performance with a predefined reference value, such as the ratio between supply and demand during a critical event or between the available energy and the required energy.
Another quantification type are \textit{score-based} resiliencemetrics. \textit{Score-based} resilience metrics rely on an qualitative or semi-qualitative assessment according to certain criteria, using e.g. a 5-point scale (from "very good" to "very bad").
Some metrics can also be composed of multiple weighted metrics - these will be referred to as \textit{composite metrics}. This approach is recommended by Hollnagel for assessing the resilience of systems in general, as he disputes that resilience is a quantity which can be captured by a single measurement ~\cite{Hollnagel.2011}.
\subsection{Metrics According to Their Scope}
\label{subsec:scope}
The key advantage of using metrics is to have a relative assessment of a certain property of a system - either with regard to its own states or with regard to other systems. With the help of resilience metrics, specifically, it should be possible to distinguish whether a new state of the system is more resilient than an older one, but also whether one system is more resilient than another. Accordingly, metrics are classified with regard to whether they are evaluated \textit{(i)} for different states of one system for comparing the resilience of those two states, and \textit{(ii)} for different systems for comparing the resilience of those two systems.

Metrics also differ in their generality with regard to critical events. By definition, a system is resilient independent of a critical event, i.e. the resilient system definition from Section \ref{subsec:overview} should hold for all reasonable critical events. Accordingly, metrics are classified with regard to whether they are evaluated in view of critical events affecting the system as well as \textit{which} and \textit{how many} different types of critical events are considered.

\section{Literature Search Protocol}
\label{sec:Search}

\begin{figure}[!htbp]
    \centering
    \includegraphics[width=12cm]{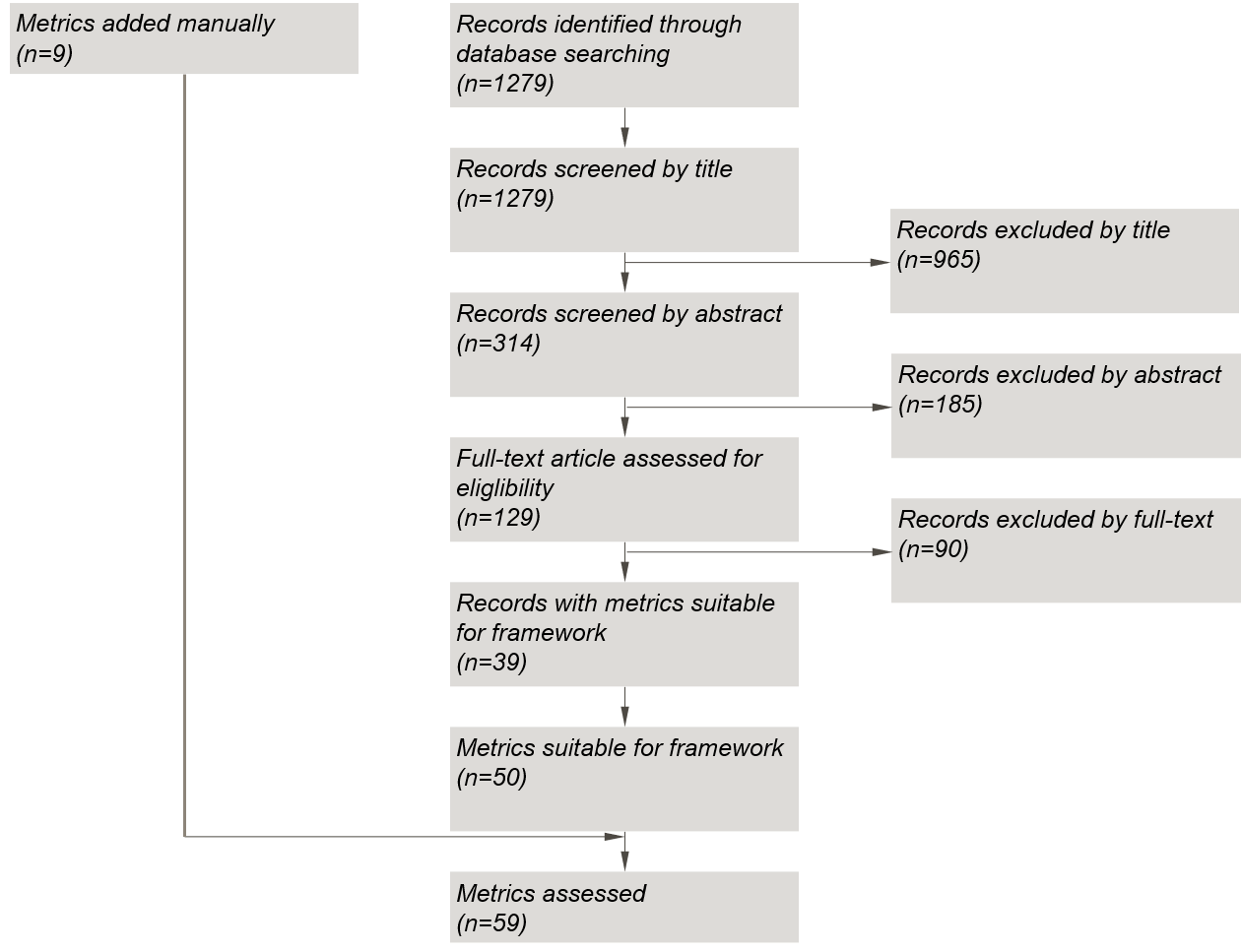}
    \caption{Flowchart representing the step-wise filtering process according to the PRISMA guidelines.}
    \label{fig:PRISMA}
\end{figure}

The categorisation of currently existing resilience metrics using the framework presented above is based on a systematic search. Following the guidelines developed under the PRISMA concept (Preferred Reporting Items for Systematic Reviews and Meta-Analyses), the following query in the Web-of-Science database was performed on March 10, 2022:

\textit{resilien* AND (metric* OR indicator* OR quantitative* OR index OR indices) AND water AND (distribution* OR supply OR network* OR infrastructure*)}. Only papers written in English were included in the study. The workflow is shown in Figure \ref{fig:PRISMA}.

\begin{figure}[!ht]
    \centering
    \scalebox{0.7}{\input{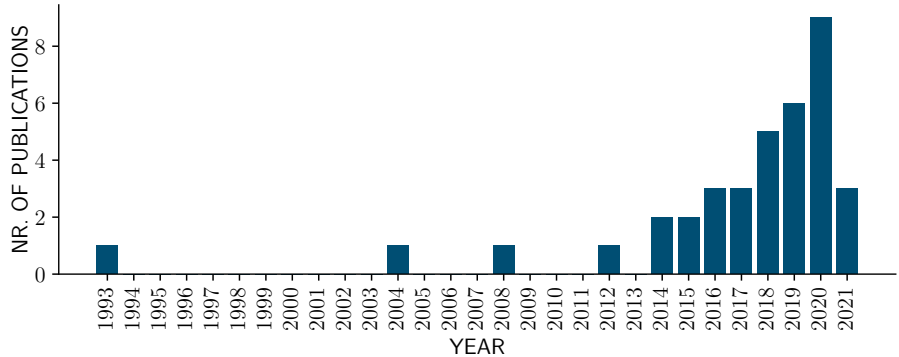}}
    \caption{A histogram of papers found in the literature search based on year of publication. After 2014, an increase in the number of publications is noticeable.}
    \label{fig:publication}
\end{figure}

The initial search led to 1279 records. The titles of these publications were screened manually, after which  965 papers were filtered out, yielding 314 records.
Subsequently, the abstracts underwent thorough screening. Whenever an abstract stated that a newly developed or adapted resilience metric was proposed or discussed within the paper, the paper was considered for further reading. Through this process, 185 papers were filtered out, leading to 129 papers.

Finally, full-text screens of all of the 129 papers were performed in order to assess whether the paper contains metrics that are presented as resilience metric, and whether the metrics are newly introduced or adopted from previous work. In this step, 90 papers were filtered out, resulting in 39 papers that contained suitable metrics. Most of the papers were published after 2014 as can be seen in the histogram in Figure \ref{fig:publication}. Since some papers contained more than one metric, 50 resilience metrics were found through this search. Since several resilience metrics known to the authors were not captured by the search query, they were added manually (9 metrics). This leads to a total of 59 resilience metrics. The initial list of publications and all filtering steps can be followed with the help of the table provided in the supplementary material, c.f. Section \ref{sec:data}.

The resilience metrics were categorised using the framework introduced in Section \ref{sec:Framework} and subsequently used to answer the research questions from Section \ref{sec:Introduction}.



\section{Results - Analysis of the Literature Review}
\label{sec:Results}
\subsection{How do existing metrics assess resilience?}


Most metrics (46; 78\%) assess resilience based on the performance of the system, i.e. on system output (such as delivered head or volume flow). 4 (7\%) metrics are score-based, evaluating resilience of a system using a score system, and 9 (15\%) metrics are based on approaches from graph theory. There are 15 composite metrics that combine multiple metrics using normalisation and weighting factors. Composite metrics can be composed of metrics of one quantification type or combine multiple ones (e.g. graph-theoretical and score-based).

From a temporal perspective, the review shows that there are 34 (58\%) time-independent resilience metrics and 25 (42\%) time-dependent resilience metrics.

All graph-theoretical metrics are time-independent. The point-based metrics are predominantly time-independent, and the performance-based metrics are about 50/50 split between time-dependent and time-independent.

In total, 35 metrics use normalisation to a certain interval, most commonly $[0, 1]$. In most cases, the optimal value is 1. In other cases, there is no upper or lower bound to resilience.

\subsection{To what extent do the existing metrics assess resilience with regard to the functions and properties of resilient systems?}

\begin{figure}[!ht]
    \centering
    \includegraphics[width =12cm]{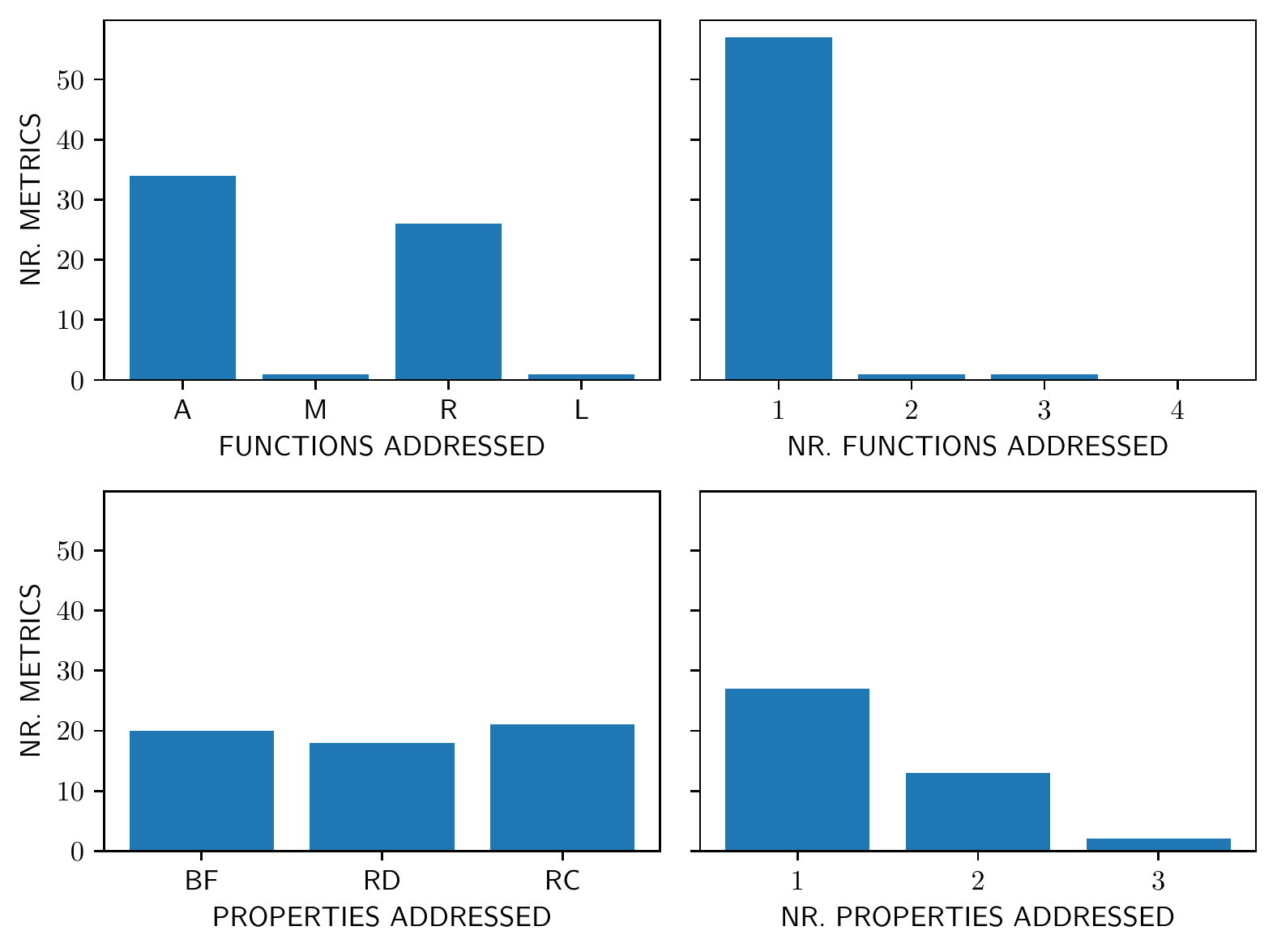}
    \caption{Metrics by resilient system functions and properties they address. Most metrics only assess either anticipating or reacting. Baseline functionality (bf), redundancy (red) and recovery (rec) are addressed by about 30\% of the metrics each. }
    \label{fig:metrics_by_functions}
\end{figure}

There is a strong tendency to address anticipation (34; 58\%) and reacting (26; 44\%) rather than monitoring (1; 2\%) and learning (1; 2\%), even though monitoring and learning are considered vital resilience functions. 

The vast majority of metrics assesses only one function (57; 97\%). Only a single metric assesses 2 and 3 functions each; no metrics assess all four functions. A thorough assessment of resilience, considering all resilience functions, is thus missing.

Properties of resilient systems - baseline functionality, redundancy and recovery are addressed by 20, 18 and 21 metrics, respectively (about 30\%). Similarly to the functions, most metrics that address properties of resilient systems only address one of them (27; 46\%). 13 (22\%) metrics consider 2 properties, and two considers all three.

\begin{figure}[!ht]
    \centering
    \scalebox{0.75}{\input{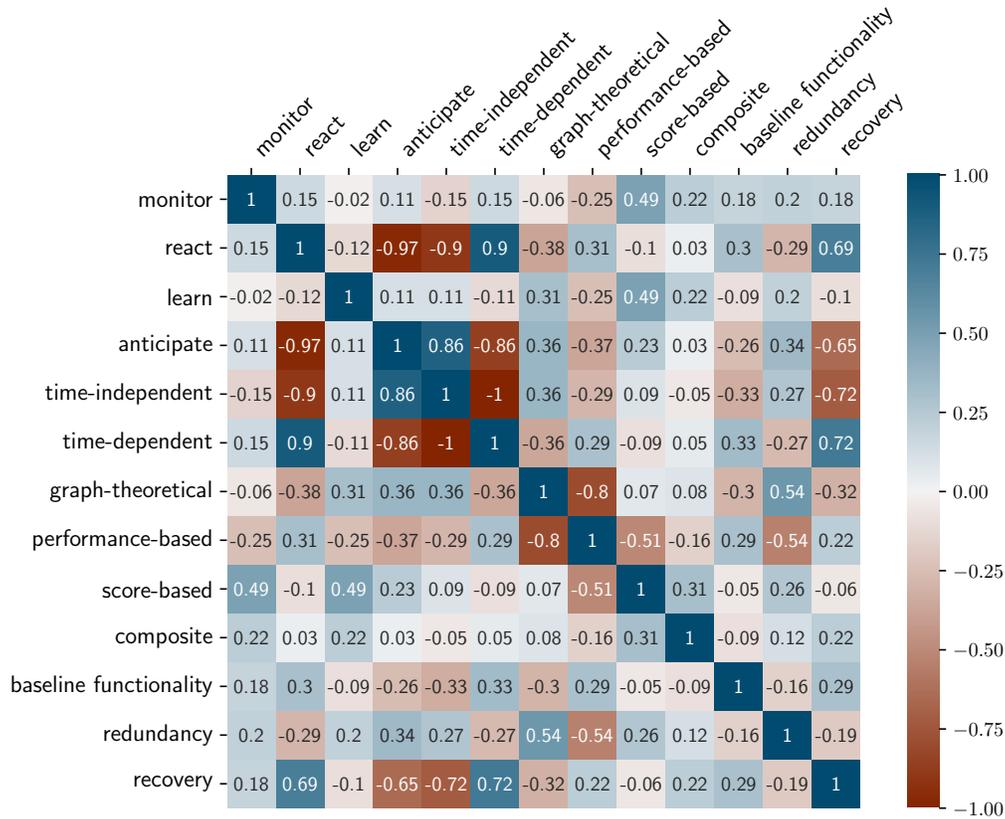}}
    \caption{Pearson correlation matrix between the data categories "monitor",	"react",	"learn",	"anticipate", "time-independent",	"time-dependent","graph-theoretical","performance-based","score-based",	"composite",	 "baseline functionality","redundancy",	"recovery". Positive values (blue) and negative values (red) suggest positive and negative correlation, respectively.}
    \label{fig:correlation}
\end{figure}

To assess the relationship between the characteristics of the metrics and those of the systems, as well as between the functions and properties addressed, Pearson correlation coefficient was calculated for selected data categories, see  correlation matrix in Figure \ref{fig:correlation}. The results show that there is a strong positive correlation between time-dependence and the react function, and time-independence and the anticipate function, meaning that metrics that assess reaction tend to be time-dependent, while metrics that assess anticipation tend to be time-independent. Assessing anticipation and being graph-theoretical, as well as assessing reaction and being performance-based correlate moderately. Assessing the recovery property correlates strongly with being time-dependent. The property redundancy correlates moderately with being graph-theoretical.

\subsection{How general are the existing resilience metrics with regard to different
critical events?}

\begin{figure}[!ht]
    \centering
    \includegraphics[width =12cm, trim={2cm 0cm 0cm 0cm}]{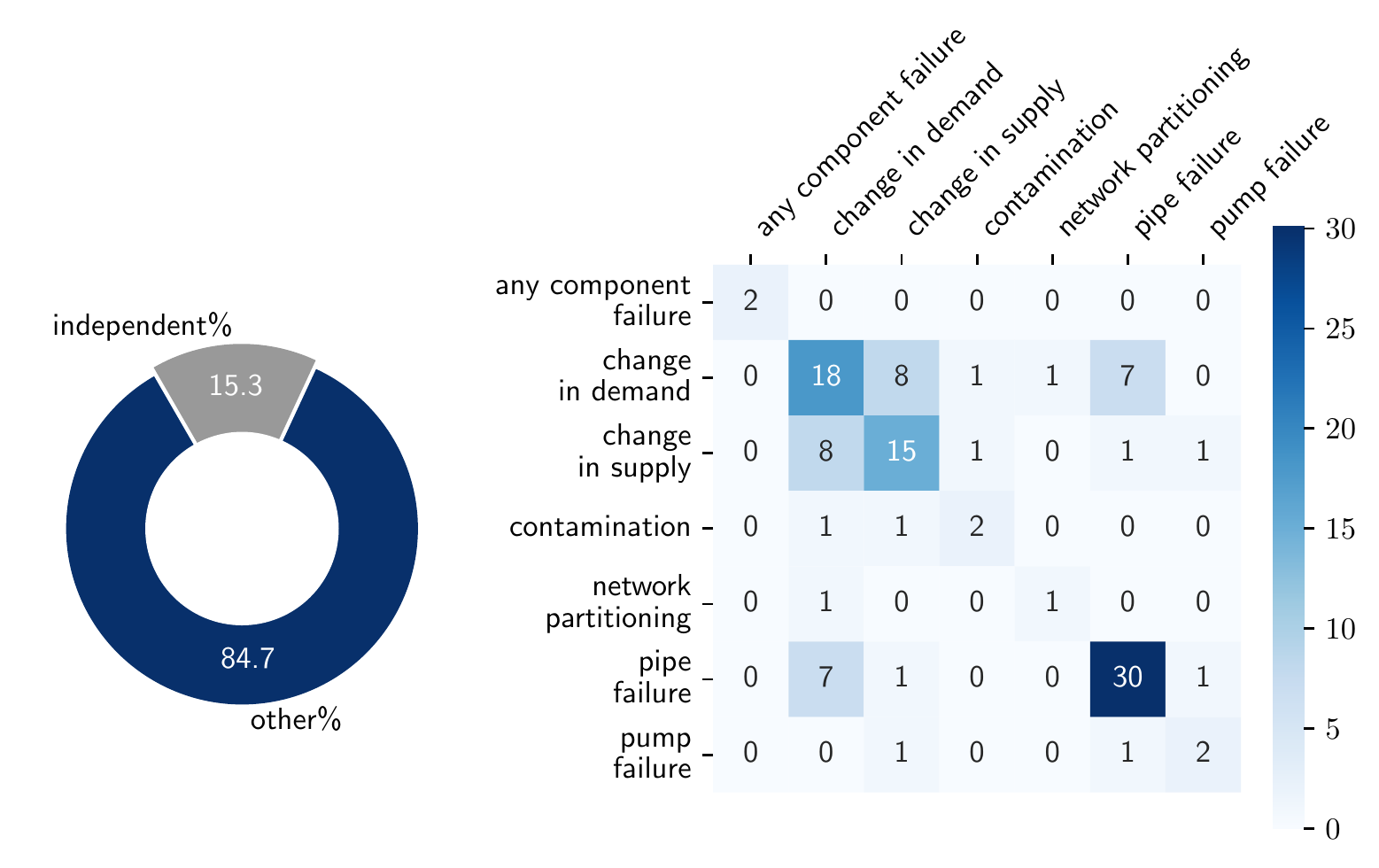}
    \caption{Left: Metrics based on whether they are independent of critical events. Only 15.3 \% of metrics are independent. Right: Numbers of metrics based on which critical events they can capture. Most metrics can only capture one; combinations of change in supply and change in demand, as well as of pipe failure and change in demand are relatively common.}
    \label{fig:crit_events}
\end{figure}

According to the resilience understanding in Section \ref{sec:Resilience}, a resilient system can keep its minimum functionality in any (reasonable) critical event. Hence, the metrics should aim for  independence from the type of critical events.

The results of this study show that only about 15\% of the reviewed metrics are independent of critical events (Figure \ref{fig:crit_events}). Figure \ref{fig:crit_events} also shows that most metrics focus on a specific subset of critical events, and most commonly only on one (pipe failure - 21, change in demand - 2, change in supply - 5). Two metrics consider "any component failure", which is a relatively general category that can include pump failure, pipe failure, valve failure and other component failures.
A commonly occurring combination is between change in demand and change in supply (8), as well as between pipe failure and change in demand (7). Only a single metric combines three critical events.

With this limited view, it can be argued that the metrics assess robustness of the system with regard to pipe failure or change in demand/supply, rather than its resilience.


11 metrics from this review have been used to assess different systems and make a comparison between them. The networks used for resilience assessment are also stated in the dataset linked in Section \ref{sec:data}: except the benchmark network Net3 which was used in 3 cases, no pattern with regard to the usage of networks can be observed.

\subsection{Clustering Reviewed Metrics}
\label{sec:clustering}

\begin{figure}[!hp]
    \centering
    \scalebox{1.0}{\input{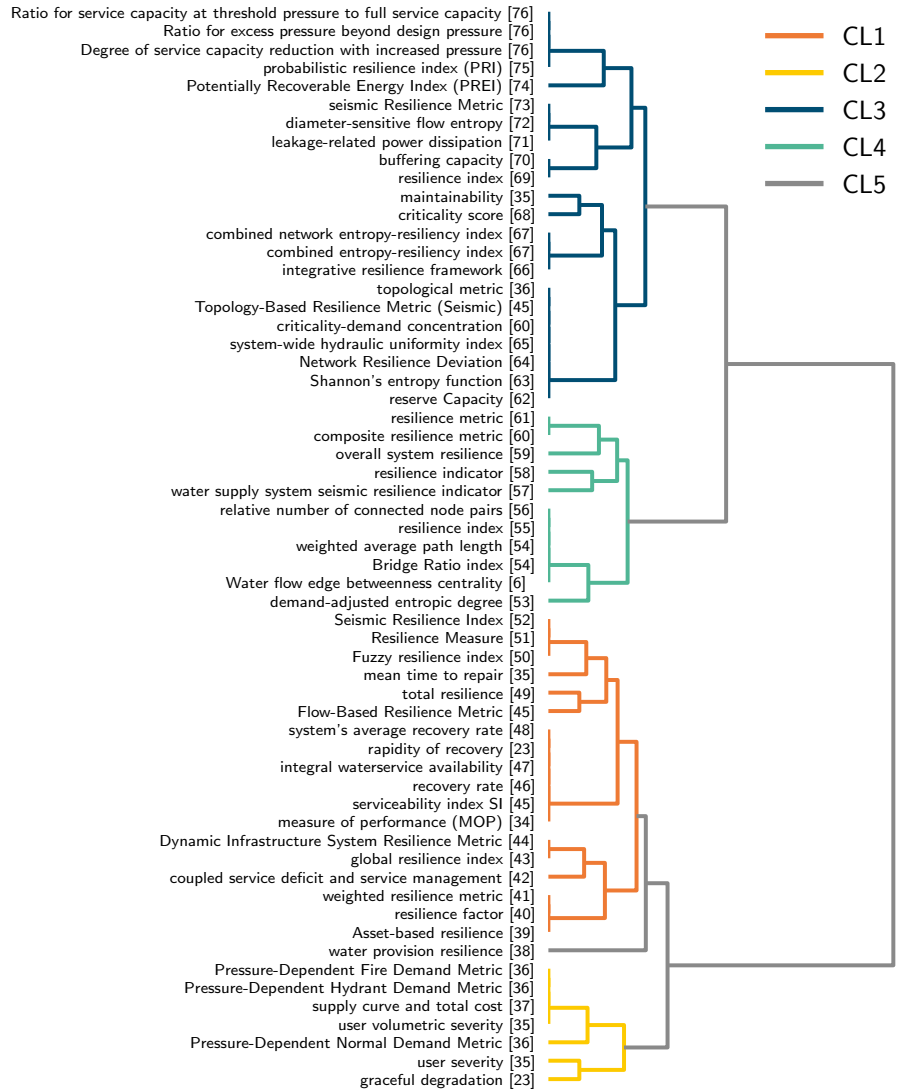}}
    \caption{Dendrogram showing all reviewed resilience metrics grouped into 5 clusters (CL1-5). For assignment of metrics to clusters in text form, consult Table \ref{tab:categorisation}.}
    \label{fig:dendrogram}
\end{figure}

Hierarchical clustering has been performed on the reviewed metrics along the categories "systems functions addressed", "system properties addressed", "dependence on time", and "quantification". More details on the clustering algorithm are provided in Section \ref{a:clustering}. The results from the distance matrix are plotted in Figure \ref{fig:dendrogram}, forming 5 clusters (CLs). The clusters can be characterised as follows:

\begin{itemize}

    \item CL1: reaction metrics considering recovery
    \item CL2: reaction metrics not considering recovery
    \item CL3: performance-based anticipation metrics
    \item CL4: non-performance-based anticipation metrics
    \item CL5: score-based resilience metrics considering all properties
\end{itemize}

\subsection{Categorisation of Selected Resilience Metrics for WDS}

Below, selected metrics characteristic for each of the clusters specified in Section \ref{sec:clustering} are presented. The full list of reviewed metrics is presented in the \ref{a:categorisation}.

\subsubsection{Reaction Metrics Considering Recovery (CL1)}
In CL1, all metrics are performance-based, assess reaction and consider recovery. They all consider recovery. Many of them also address baseline functionality.

Hashimoto et al. define the \textit{system's average recovery rate} as a measure of resilience \cite{Hashimoto.1982}. For the system output $X_t$ at time $t$ which can be in a satisfactory state $S$ or failure state $F$, the metric can be expressed as follows:

\begin{equation}
    \gamma = \frac{P(X_t \in S  \: \mathrm{and} \:  X_{t+1} \in F)}{P(X_t \in F)} = \frac{\varrho}{1 - \alpha},
\end{equation}

where $\varrho$ denotes the probability $P$ of the system transitioning from the set $S$ in the period $t$ to the set $F$ in the period $t+1$, and $\alpha$ denotes the probability of being in a satisfactory state: $\alpha = P(X_t \in S)$ \cite{Hashimoto.1982}.

The metric is designed to aid in determining design and operating policies for WDSs \cite{Hashimoto.1982}.
In the understanding of Hashimoto et al., resilience describes "how quickly a system is likely to recover or bounce back from failure once failure has occurred" \cite{Hashimoto.1982}.
This measure assesses reaction, namely how likely the system is to to transition back to a satisfactory state after failure. Hence, it considers recovery after the failure, and also requires baseline functionality to define the satisfactory/failure state.
It is a time-dependent, performance-based metric. Hashimoto et al. do no not prescribe what quantity the system output  $X_t$ should be expressed with, but in their case study they work with volume.

System's avergage recovery rate is defined on the interval $[0,1]$ with 1 being the optimum value.
The use of the metric is illustrated on a water reservoir with seasonal changes in demand and supply.

Zhuang et al. define their resilience metric \textit{integral water service availability} as "the percentage of water supplied to customers over a system failure period" \cite{Zhuang.2013}. It can be expressed as the ratio of delivered flowrate $Q$ (supply) to required flowrate $Q^*$ (demand) over the selected period of time when the critical event occurred \cite{Zhuang.2013}. At system scale, it is formulated as

\begin{equation}\label{zhuang_metric}
     R_\mathrm{sys} = \frac{\sum_{t=1}^{T} \sum_{i=1}^{N}Q_{i,t}}{\sum_{t=1}^{T} \sum_{i=1}^{N}Q_{i,t}^*}
\end{equation}

with $T$ being the time duration under system failure and $N$ the number of nodes \cite{Zhuang.2013}.
Using Monte-Carlo simulations, Zhuang et al. aim to assess the performance of the studied networks under various conditions. They aim to investigate what the critical factors affecting system resilience are. Moreover, they demonstrate how the expected costs for improving the WDS resilience can be determined.
Zhuang et al. understand resilience as "the ability to recover from a failure to a satisfactory state" \cite{Zhuang.2013}. They consider the duration of recovery an important aspect of resilience. Considering availability \eqref{zhuang_metric} to be a resilience metric, they argue that it also provides an insight into the intensity of the critical event.
This metric is a typical time-dependent resilience metric. It considers the time interval after the critical event, reflecting the focus on recovery.
As such, it is capable of assessing the reacting function of the system. The authors use the metric to study the WDS performance when the current mode of functioning is adjusted, i.e. under operator intervention or under adaptive pump operation. Baseline functionality is reflected in the denominator (required volume flow $Q^*$).
It is a performance-based metric defined on an interval between 0 and 1 with 1 being the optimal value at which the demand can be satisfied during the entire time duration under system failure. Performance of the system is measured using volume flowrate $Q$.
The metric was applied to a medium-sized example network representing a primarily residential community. Comparisons are made between different reaction strategies.
Zhuang et al. use the metric to study WDS resilience under randomly generated changes in demand and pipe failures.

Farahmandfar and Piratla define a \textit{flow-based resilience metric} ~\cite{Farahmandfar.2018} derived from Todini~\cite{EzioTodini.2000}:
\begin{equation}
    \mathrm{FR} = \frac{\sum_{t=1}^{td} \sum_{i=1}^{N_n}
                        \left[\left(\sum_{j=1}^{N_i}(1-P_{fj})\right)
                        q_{i,t}^*(h_{i,t}-h_{i,t}^*)\right]
                        }{4 \times \sum_{t=1}^{td} \sum_{i=1}^{N_n} q_{i,t}^* h_{i,t}^*}.
\end{equation}
Here, $q_{i,t}^*$ is the design demand at node $i$ in time step $t$, $h_{i,t}^*$ is the minimum required total head at node $i$ in time step $t$, and $h_{i,t}$ is the actual total head at node $i$ in time step $t$.
The factor $(1-P_f)$ represents pipe reliability using pipe fragility $P_f$, which is computed for each pipe $j$ as
\begin{equation}
    P_{fj} = 1-\exp(-\mathrm{RR}_j\cdot L_j),
\end{equation}
with repair rate $\mathrm{RR}_j$ of pipeline $j$ and length $L_j$ of pipeline $j$.
The quantities are summed over the number of time steps in the demand pattern $td$, the total number of nodes in the WDS $N_n$, and the node degree of node $i$ $N_i$.
The metric is used for making decisions in rehabilitation schemes with the objective of enhancing resilience within budgetary constraints.
Farahmandfar and Piratla state that resilience refers to the ability of WDSs to "withstand stresses, mitigate failures, minimise consequences, and recover quickly in the face of abnormalities such as earthquakes"~\cite[p.1]{Farahmandfar.2018}.
The metric assesses the reaction of the WDS. Similarly to the resilience index of Todini which it is based on, it considers the properties baseline functionality and redundancy (measuring the surrogate energy of the system). Summing the values over time, however, makes it possible to also account for recovery.
It is a time-dependent metric.
It is a  performance-based metric, with performance being expressed in terms of power proportional to the product of the volume flow and head, $Qh$.
The metric is usually constrained to the interval $[0,1]$ with $1$ being its optimum value.
The metric is evaluated for a single network and considers the scenario of pipe failures due to seismic events.

\subsubsection{Reaction Metrics Not Considering Recovery (CL2)}
In CL2, all metrics are performance based and assess reaction. None of them consider recovery. They are predominantly time-dependent, and a few consider baseline functionality.

Huizar et al. propose a resilience metric called \textit{user  severity}, defined as "the minimum ratio of supply to demand, or minimum functionality, during the analysis period"  \cite{Huizar.2018}. For the $i$-th user, it is defined as

\begin{equation}
     \mathrm{US}_{i,t} = \min_{T_0 \leq t \leq T}\{ f_{i,t}\},
\end{equation}

where $T_0$ and $T$ are the beginning and the end of the analysis period and $f_{i,t} = S_{i,t}/D_{i,t}$ is the user functionality, defined as the ratio of supply $S$ to demand $D$ at time $t$ \cite{Huizar.2018}.

The metric was developed along other metrics for the purpose of measuring water system security.
Huizar et al. understand resilience as "the ability to mitigate and recover from failure" \cite{Huizar.2018}.
User severity assess the reaction of the system to a failure.
It is time-independent and performance-based; the performance is expressed in terms of volume supplied.
For non-zero demand $D$, user severity can attain values between 0 and 1, with 1 being the optimal value.
The metric is sensitive to changes in supply and demand.

\subsubsection{Performance-Based Anticipation Metrics (CL3)}
In CL3, all metrics assess anticipation and are performance-based. They are predominantly time-independent and consider baseline functionality or redundancy, in a few cases also recovery.

Todini's resilience index for looped water distribution networks \cite{EzioTodini.2000}, which is one of the most commonly used resilience metrics for WDSs, belongs to this cluster. It was formulated as

\begin{equation}
    I_\mathrm{r} = \frac{\sum^{n_\mathrm{n}}_{i=1}q_i^*(h_i - h_i^*) }{\sum^{n_\mathrm{r}}_{k=1}Q_k H_k + \sum^{n_\mathrm{p}}_{j=1}(P_j/\gamma) - \sum^{n_\mathrm{n}}_{i=1}q_i^*h_i^*},
\end{equation}

with $q_i^*$ and $h_i^*$ being the design demand and head required at the node $i$, $h_i$ being the available head at the node $i$, $Q_k$ and $H_k$ being the flow from and total head in the $k$-th reservoir, $P_j$ is the power introduced to the network by the $j$-th pump, $\gamma$ being the specific weight of water, and $n_\mathrm{n}$, $n_\mathrm{r}$, $n_\mathrm{p}$ being the number of nodes, reservoirs and pumps, respectively \cite{EzioTodini.2000}.

Todini understands resilience as "capability of overcoming stress or failure conditions" or "the capability to allow to overcome local failures and to guarantee the distribution of water to users"\cite{EzioTodini.2000}.

The resilience index is a time-independent resilience metric. It is perfor-mance-based, comparing the required power with the available power in the WDS. It can have values between 0 and 1, with 1 being the optimum value.
Todini illustrates the use of the index on optimisation problems with three simplified looped networks with the aim of minimum cost design.
He uses the resilience index in the design phase in order to develop a heuristic optimisation approach to arrive at a Pareto set of solutions in the cost vs. resilience space.

While Todini makes comparisons in the resilience index for a specific WDS, he does not make comparisons between systems.
The resilience index is independent of critical events.
It assesses the anticipating function of the system. It is not necessary for the critical event to occur in the analysis or in the real world in order to be able to assess it.


Altherr et al. bring the \textit{buffering capacity} into resilience engineering \cite{Pelz.2021b}. This metric was first described by Woods as “the size or kinds of disruptions the system can absorb or adapt to without a fundamental breakdown in performance or in the system’s structure” \cite{Woods.2017}. Altherr al. defined buffering capacity as "a measure for the amount of structural change after which the fulfillment of a predetermined required minimum of functional performance can still be guaranteed" \cite{Altherr.2018}. In WDSs, buffering capacity can be expressed using discrete values - the number $k$ of components that can fail while the  minimum of functional performance can still be guaranteed. The system is then called \textit{k-resilient}.

\subsubsection{Non-Performance-Based Anticipation Metrics (CL4)}
In CL4, all metrics are time-independent and assess anticipation. They are graph-theoretical or score-based. Composite anticipation metrics also belong to CL4. Most of these metrics address redundancy.

Herrera et al. base their resilience index on a common graph-theoretical algorithm, K-shortest paths \cite{Eppstein.1998}. The index is also extended to WDSs sectorised into district metered areas. The measure of resilience is first computed for each node by determining the K-shortest paths between the node and each source. To account for hydraulics, the paths are weighted by energy loss associated with the flow resistance along the path. The resilience index of Herrera et al. for a node $i$ is mathematically defined as

\begin{equation}
    I(i) = \sum_{s=1}^{S}\left(\frac{1}{K}\sum_{k=1}^K\frac{1}{r(k,s)}\right)
\end{equation}

with $S$ being the total number of sources, $K$ the number of shortest paths  and $r(k, s)$ being the measure of the energy loss for the path $k$ to source $s$ \cite{Herrera.2016}. It can be expressed for example as follows:

\begin{equation}
    r(k) = \sum_{m=1}^M f(m)\frac{L_m}{D_m}
\end{equation}

with $M$ being the number of pipes on the path $k$, $f$ the friction factor and $L$ and $D$ the pipe length and diameter, respectively \cite{Herrera.2016}. As Lorenz and Pelz show, the index can be additionally weighted by relative node demand $q/Q$ where $q$ is the node demand and $Q$ the total demand in the network \cite{Lorenz.2020}.

For a district metered area, Herrera et al. propose aggregating the resilience indices of each node $j$ into a single resilience index for all $n$ nodes using the trimmed mean \cite{LuisAngelGarciaEscudero.2003}:

\begin{equation}
    I^* = \sum_{j=1}^{n^*} \frac{I(j)}{n^*}
\end{equation}
in which nodes of very high or very low values are discarded before computing the mean ($n^*<n$) \cite{Herrera.2016}.
The purpose of the work of Herrera et al. is to develop a resilience assessment framework. The index is shown to be consistent with other alternative approaches.
Herrera et al. understand resilience as "the ability of a system to maintain and adapt its operational performance in the face of failures and other adverse conditions" \cite{Herrera.2016}.
The resilience index of Herrera et al. addresses anticipation. It is a measure of the system's expected behaviour during a critical event. It quantifies redundancy in connectivity and supply \cite{Herrera.2016}. It is not capable of considering recovery.
It is a time-independent metric that only considers topology of a network and hydraulic properties.

Herrera et al. validate the metric on the C-Town network~\cite{Ostfeld.2016} and they use it to analyse the resilience of two networks with 4820 and 106,115 nodes, respectively. Comparisons are made between the resilience of various DMAs, not between the networks. The considered disruption event is pipe failures.

Balaei et al. developed a framework for assessing resilience, leading to the \textit{water supply system resilience indicator} that aggregates several weighted and scaled metrics:

\begin{equation}
    R = \frac{1}{\sum_{j=1}^N w_j} \sum_{j=1}^N w_j i_j^2,
\end{equation}

where the weights are denoted by $w$ and the indicators by $i$ for $N$ indicators in total \cite{Balaei.2018}. The indicators are scaled by the respective maximum value. The indicators are "operational representations of serviceability, quality, or a characteristic of a system" \cite{Balaei.2018} that satisfy the criteria of validity, sensitivity, objectivity and simplicity \cite{Balaei.2018}. A specific set of indicators must be chosen for each use case under the consideration of data availability. Examples of indicators provided in the paper are physical vulnerability, knowledge of the emergency response plan, social participation rate, GDP per capita and median household's income.
The purpose of the resilience indicator and the proposed framework is to assess seismic resilience based on data and information from past earthquakes. The framework is aimed at researchers, planners and decision makers.
The metric considers the anticipating function.
It is a time-independent metric.
It is a score-based metric.
It has been evaluated on one example without comparisons. It has a strong focus on earthquakes but as the choice of indicators has to be determined for each individual system, there is potential to adjust it to other critical events as well.

\subsubsection{Score-based Resilience Metrics Considering All Properties (CL5)}
Among the metrics found in the presented study, CL5 contains only one score-based metric that considers three system functions (monitor, react and anticipate) and all three properties. It is thus the closest to being a resilience metric.

The \textit{water provision resilience (WPR)} was proposed by Milman and Short~\cite{Milman.2008}.
Rather than giving a single equation for calculating it, WPR is an aggregate of points that the considered WDS scores in the categories supply, finances, infrastructure, service provision, water quality, and governance.
In each of these categories, there are different numbers of criteria for which a binary decision is made whether they are fulfilled or not, each fulfilled criterium yielding a point.
The sum of points gives the score of WPR.
The purpose of this metric is not, as the authors state, to "measure the adaptive capacity related to catastrophic events"~\cite[p. 756]{Milman.2008}. Instead, the focus lies on measuring "the ability of a city or water district to maintain or improve access to safe water"~\cite[p.760]{Milman.2008}.
In their understanding of resilience, the authors refer to~\cite[p.259]{Folke.2006} understanding resilience as "the capacity of the system 'to absorb disturbance and re-organise while undergoing change so as to still retain essentially the same function, structure, identity, and feedbacks'", emphasising that the definition includes the ability of the given system "to adapt to stresses and changes and to transform into more desirable states"~\cite[p.759]{Milman.2008}.
The variety of criteria included for the resilience evaluation allow the metric to cover three of the resilience functions: monitor, react and anticipate.
The properties baseline-functionality, redundancy and recovery are also considered within the criteria.
As the criteria include the development of the WDS within the following 50 years, the metric is time-dependent.
It is a score-based and composite metric.
In total, 36 criteria are included, i.e. the maximum achievable value of WPR is 36, the minimum being 0.
The metric is used by Milman and Short to assess the resilience of the WDS of three municipal areas and to compare the resilience of these.
Critical events considered in the criteria are change in demand, change in supply, and water resource contamination.

\section{Discussion}
\label{sec:Discussion}

In a systematic review of resilience metrics for WDSs, the presented results show that most metrics, regardless of what their characteristics are, only focus on a single function and/or property of resilient systems, rather than on their resilience as a whole.
The review bridges a gap in research about resilience metrics for WDSs as it provides a comprehensive framework for categorising metrics and juxtapose them with a general understanding of resilience.

Most often, the functions "anticipate" and "react" are assessed.
While generality with regard to critical events is often stressed when speaking about resilience, it is not reflected in the metrics which tend to focus on specific critical events such as pipe failure and changes in demand or supply.
Moreover, that the metric is defined on a specific interval with an optimal value suggests that the system can achieve perfect resilience. Once the system achieves it, there is no more room for improvement with regard to resilience. It is, however, questionable whether such a state is achievable for real-world networks, and whether the resilience metrics are really capable of capturing this.

Strictly speaking, the presented assessment framework shows that there is no metric among the existing metrics reviewed that can be called a \textit{resilience metric}, as no metric addresses all 4 functions of a resilient system.
This is not to say, however, that the metrics are not useful for certain purposes, even for those related to resilience assessment, or e.g. optimisation for resilience. Resilience is a complex concept that is difficult to capture by quantitative and even qualitative metrics.
Instead, the authors propose that a stronger differentiation should be made among metrics related to resilience assessment in WDSs: for example, to speak of anticipation metrics or reaction metrics rather than of resilience metrics. This will help prevent conceptual stretching of the term resilience, already criticised nowadays for being a buzzword or an umbrella term particularly difficult to work with in academia \cite{Fekete.2020, Bogardi.2019}. The presented framework can be used for this purpose.

The design of the presented framework depends strongly on the selected definition of resilience. As no scientific consensus with regard to the definition of resilience exists, the authors have selected a definition that is well-known and general enough to cover most other definitions present in literature.
The assessment of functions and properties by metrics has been a challenging task during the review that is necessarily prone to a certain amount of subjectivity. By providing both the data and the code used for the analysis (Sec. \ref{sec:data}), the authors hope to lay ground for a discussion of the framework and resilience understanding in the domain of WDS.

While difficult to capture by metrics, the authors are of the opinion that the understanding of resilience should not be limited in order to make it easier to quantify, but rather that new metrics should be developed in order to improve its quantifiability. Especially the functions "learn" and "monitor" are largely ignored by the existing metrics for WDS. Existing frameworks such as the Resilience Analysis Grid \cite{Hollnagel.2011} or water provision resilience \cite{Milman.2008} can be used as a guideline; while having a thorough resilience understanding, these frameworks, however, lack quantitative metrics and are thus currently difficult to implement in studies commonly performed in the field of resilience engineering of WDSs, such as optimisation problems or Monte Carlo simulations.

The presented results also prepare ground for further research in the domain of WDS resilience. A big challenge remains to systematically incorporate climate change effects into resilience metrics for WDS, as climate change is/ is the cause of critical events that affect water distribution. In some cases, it will be necessary to extend the system boundary of WDS to include water resource management and/or other infrastructure that can be used for delivering water to citizens, such as the transport network.
Moreover, like other disciplines \cite{Fekete.2020, Cai.2020, Canizares.2021}, WDS research should also take a critical look on resilience, evaluating the weaknesses and strengths of the concept and reflect these in the metrics.

\section{Conclusion}

The presented publication assessed the alignment between a general understanding of resilience in water distribution systems and the metrics used for their resilience assessment. For this purpose, a systematic review of resilience metrics for WDSs was performed, showing that:
\begin{itemize}
    \item most metrics are performance-based rather than graph-theoretical or score-based, and time-independent rather than time-dependent (RQ1)
    \item most metrics, regardless of what their characteristics are, only focus on a single function and/or property of resilient systems, rather than on their resilience as a whole (RQ2)
    \item most metrics focus on a specific set of critical events, resulting in a lack of generality inherent to the understanding of resilience (RQ3)
\end{itemize}

To summarise and answer the title question, the results show that resilience metrics do not really assess resilience, but rather specific functions and properties of systems which can make them resilient. To prevent further conceptual stretching of the term resilience, the authors propose that a stronger differentiation is made among metrics related to resilience assessment in WDSs: for example, to speak of anticipation metrics or reaction metrics rather than of resilience metrics.

\section{Data and Software Availability}\label{sec:data}
The data for this study (a table with categorisation of all reviewed metrics as well as a table with the literature search procedure) is available under \url{https://tudatalib.ulb.tu-darmstadt.de/handle/tudatalib/3900}.

The corresponding code in the form of Jupyter notebooks is available under \url{https://tudatalib.ulb.tu-darmstadt.de/handle/tudatalib/3901}.

\section{Author Contributions}
\textbf{Michaela Le\v{s}táková}: Conceptualization, Methodology, Software, Formal analysis, Investigation, Writing - Original Draft, Writing - Review \& Editing, Visualization, Data Curation, Project administration; \textbf{Kevin T. Logan}: Conceptualization, Methodology, Investigation, Writing - Original Draft, Writing - Review \& Editing, Visualization, Data Curation, Project administration; \textbf{Imke-Sophie Rehm}: Conceptualization, Methodology, Investigation, Writing - Original Draft, Writing - Review \& Editing, Project administration; \textbf{John Friesen}: Conceptualization, Methodology, Investigation, Resources, Writing - Original Draft, Writing - Review \& Editing, Project administration Supervision; \textbf{Peter F. Pelz}: Funding Acquisition, Supervision

\section{Acknowledgements}
This work has been funded by the LOEWE initiative (Hesse, Germany) within the emergenCITY center, by the LOEWE exploration project "Uniform detection and modeling of slums to determine infrastructure needs" as well as by the KSB Stiftung Stuttgart, Germany within the project “Antizipation von Wasserbedarfsszenarien für die Städte der Zukunft”.

The authors would like to thank Yali Wu for her help with performing the correlation analysis of the reviewed resilience metrics and Katharina Henn for the literature search.

\newpage
\appendix
\section{Clustering}\label{a:clustering}
The hierarchical clustering was performed in Python utilising the methods \texttt{scipy.cluster.hierarchy.linkage} (method: \texttt{'ward'}, \newline metric: \texttt{'Euclidian'}) and \texttt{scipy.cluster.hierarchy.fcluster} (number of clusters: 5, criterion: \texttt{'maxclust'}).

The dendrogram was created with the method \texttt{dendrogram} \newline from \texttt{scipy.cluster.hierarchy}.

The code including an Anaconda environment file with all necessary Python packages is available in the Jupyter Notebook provided in the Section \ref{sec:data}.

\section{Categorisation of Metrics}\label{a:categorisation}
\begingroup


\fontsize{8}{11}\selectfont

\begin{longtblr}[
   caption = {Categorisation of the existing resilience metrics. M: monitor, R: react, L: learn, A: anticipate, TI: time-independent, TD: time-dependent, GT: graph-theoretical, PB: performance-based, SB: score-based, CM: composite, BF: baseline functionality, RD: redundancy, RC: recovery, CL: cluster},
   entry = 1,
   label = tab:categorisation,
 ]{
   colspec = {p{4cm}p{0.1cm}p{0.1cm}p{0.1cm}p{0.1cm}p{0.25cm}p{0.25cm}p{0.25cm}p{0.25cm}p{0.25cm}p{0.25cm}p{0.25cm}p{0.25cm}p{0.25cm}p{0.25cm}p{0.25cm}},
   rowhead = 1,
  row{odd} = {white},
  row{even} = {gray!25},
 }

\hline
           metric & M & R & L & A & TI & TD & GT & PB & SB & CM & BF & RD & RC &  CL \\
\hline
measure of performance (MOP) \cite{Cassottana.2021} &   & 1 &   &   &    &  1 &    &  1 &             &    &  1 &    &  1 &   1 \\
                                 Dynamic Infrastructure System Resilience Metric \cite{Kong.2019} &   & 1 &   &   &    &  1 &    &  1 &             &  1 &    &    &  1 &   1 \\
                                                                total resilience \cite{Zhao.2017} &   & 1 &   &   &    &  1 &    &  1 &             &    &    &  1 &  1 &   1 \\
                                                 serviceability index SI \cite{Farahmandfar.2018} &   & 1 &   &   &    &  1 &    &  1 &             &    &  1 &    &  1 &   1 \\
                                            Flow-Based Resilience Metric \cite{Farahmandfar.2018} &   & 1 &   &   &    &  1 &    &  1 &             &    &  1 &  1 &  1 &   1 \\
                               coupled service deficit and service management \cite{Krueger.2019} &   & 1 &   &   &    &  1 &    &  1 &             &  1 &    &  1 &  1 &   1 \\
                                                     weighted resilience metric \cite{Zhang.2020} &   & 1 &   &   &    &  1 &    &  1 &             &  1 &  1 &    &  1 &   1 \\
                                                       Resilience Measure \cite{Nasrazadani.2020} &   & 1 &   &   &    &  1 &    &  1 &             &    &    &    &  1 &   1 \\
                                                        Seismic Resilience Index \cite{Liu.2020b} &   & 1 &   &   &    &  1 &    &  1 &             &    &    &    &  1 &   1 \\
                                                            resilience factor \cite{Cubillo.2019} &   & 1 &   &   &    &  1 &    &  1 &             &  1 &  1 &    &  1 &   1 \\
                                                   global resilience index \cite{Cimellaro.2016b} &   & 1 &   &   &    &  1 &    &  1 &             &  1 &    &    &  1 &   1 \\
                                                                    recovery rate \cite{Ren.2020} &   & 1 &   &   &    &  1 &    &  1 &             &    &  1 &    &  1 &   1 \\
                                                     Fuzzy resilience index \cite{ElBaroudy.2004} &   & 1 &   &   &    &  1 &    &  1 &             &    &    &    &  1 &   1 \\
                                                           mean time to repair \cite{Huizar.2018} &   & 1 &   &   &    &  1 &    &    &             &    &    &    &  1 &   1 \\
                                            integral waterservice availability \cite{Zhuang.2013} &   & 1 &   &   &    &  1 &    &  1 &             &    &  1 &    &  1 &   1 \\
                                             system's average recovery rate \cite{Hashimoto.1982} &   & 1 &   &   &    &  1 &    &  1 &             &    &  1 &    &  1 &   1 \\
                                                           rapidity of recovery \cite{Pelz.2021b} &   & 1 &   &   &    &  1 &    &  1 &             &    &  1 &    &  1 &   1 \\
                                                         Asset-based resilience \cite{Izadi.2020} &   & 1 &   &   &    &  1 &    &  1 &             &  1 &  1 &    &  1 &   1 \\
                                              supply curve and total cost \cite{Dubaniowski.2021} &   & 1 &   &   &    &  1 &    &  1 &             &    &    &    &    &   2 \\
                                                                 user severity \cite{Huizar.2018} &   & 1 &   &   &  1 &    &    &  1 &             &    &    &    &    &   2 \\
                                                      user volumetric severity \cite{Huizar.2018} &   & 1 &   &   &    &  1 &    &  1 &             &    &    &    &    &   2 \\
                             Pressure-Dependent Fire Demand Metric \cite{HernandezHernandez.2021} &   & 1 &   &   &    &  1 &    &  1 &             &    &    &    &    &   2 \\
                                                           graceful degradation \cite{Pelz.2021b} &   & 1 &   &   &  1 &    &    &  1 &             &    &  1 &    &    &   2 \\
                           Pressure-Dependent Normal Demand Metric \cite{HernandezHernandez.2021} &   & 1 &   &   &    &  1 &    &  1 &             &    &  1 &    &    &   2 \\
                          Pressure-Dependent Hydrant Demand Metric \cite{HernandezHernandez.2021} &   & 1 &   &   &    &  1 &    &  1 &             &    &    &    &    &   2 \\
                                            leakage-related power dissipation \cite{Creaco.2016b} &   &   &   & 1 &  1 &    &    &  1 &             &    &    &  1 &    &   3 \\
                         Ratio for excess pressure beyond design pressure \cite{Amarasinghe.2016} &   &   &   & 1 &  1 &    &    &  1 &             &    &  1 &    &    &   3 \\
                                                 Shannon's entropy function \cite{TANYIMBOH.1993} &   &   &   & 1 &  1 &    &    &  1 &             &    &    &    &    &   3 \\
                                                 Network Resilience Deviation \cite{DiNardo.2015} &   &   &   & 1 &  1 &    &    &  1 &             &    &    &    &    &   3 \\
                                         system-wide hydraulic uniformity index \cite{Jeong.2020} &   &   &   & 1 &  1 &    &    &  1 &             &    &    &    &    &   3 \\
                                                                 criticality score \cite{He.2019} &   &   &   & 1 &  1 &    &    &  1 &             &  1 &    &    &  1 &   3 \\
Ratio for service capacity at threshold pressure to full service capacity \cite{Amarasinghe.2016} &   &   &   & 1 &  1 &    &    &  1 &             &    &  1 &    &    &   3 \\
                                            integrative resilience framework \cite{Gonzales.2017} &   &   &   & 1 &  1 &    &    &  1 &             &  1 &    &    &    &   3 \\
             Degree of service capacity reduction with increased pressure \cite{Amarasinghe.2016} &   &   &   & 1 &  1 &    &    &  1 &             &    &  1 &    &    &   3 \\
                              Topology-Based Resilience Metric (Seismic) \cite{Farahmandfar.2018} &   &   &   & 1 &  1 &    &    &  1 &             &    &    &    &    &   3 \\
                                                              reserve Capacity \cite{Wright.2015} &   &   &   & 1 &  1 &    &    &  1 &             &    &    &    &    &   3 \\
                            Potentially Recoverable Energy Index (PREI) \cite{CubidesCastro.2021} &   &   &   & 1 &    &  1 &    &  1 &             &    &  1 &    &    &   3 \\
                                    combined network entropy-resiliency index \cite{Sirsant.2020} &   &   &   & 1 &  1 &    &    &  1 &             &  1 &    &    &    &   3 \\
                                            combined entropy-resiliency index \cite{Sirsant.2020} &   &   &   & 1 &  1 &    &    &  1 &             &  1 &    &    &    &   3 \\
                                               seismic Resilience Metric \cite{Farahmandfar.2017} &   &   &   & 1 &  1 &    &    &  1 &             &    &    &  1 &    &   3 \\
                                                           buffering capacity \cite{Altherr.2018} &   &   &   & 1 &  1 &    &    &  1 &             &    &  1 &  1 &    &   3 \\
                                                  diameter-sensitive flow entropy \cite{Liu.2014} &   &   &   & 1 &  1 &    &    &  1 &             &    &    &  1 &    &   3 \\
                                                          resilience index \cite{EzioTodini.2000} &   &   &   & 1 &  1 &    &    &  1 &             &    &  1 &  1 &    &   3 \\
                                     criticality-demand concentration \cite{S.S.Ottenburger.2019} &   &   &   & 1 &  1 &    &    &  1 &             &    &    &    &    &   3 \\
                                                               maintainability \cite{Huizar.2018} &   &   &   & 1 &  1 &    &    &  1 &             &    &    &    &  1 &   3 \\
                                      probabilistic resilience index (PRI) \cite{BinMahmoud.2018} &   &   &   & 1 &  1 &    &    &  1 &             &    &  1 &    &    &   3 \\
                                                topological metric \cite{HernandezHernandez.2021} &   &   &   & 1 &  1 &    &    &  1 &             &    &    &    &    &   3 \\
                              water supply system seismic resilience indicator \cite{Balaei.2018} &   &   &   & 1 &  1 &    &    &    &           1 &  1 &    &  1 &    &   4 \\
                                                             resilience index \cite{Herrera.2016} &   &   &   & 1 &  1 &    &  1 &    &             &    &    &  1 &    &   4 \\
                                      relative number of connected node pairs \cite{Dwivedi.2014} &   &   &   & 1 &  1 &    &  1 &    &             &    &    &  1 &    &   4 \\
                                              demand-adjusted entropic degree \cite{Yazdani.2012} &   &   &   & 1 &  1 &    &  1 &    &             &    &    &    &    &   4 \\
                                                              resilience metric \cite{Assad.2019} &   &   &   & 1 &  1 &    &  1 &    &             &  1 &    &  1 &    &   4 \\
                                                           Bridge Ratio index \cite{DiNardo.2018} &   &   &   & 1 &  1 &    &  1 &    &             &    &    &  1 &    &   4 \\
                                        Water flow edge betweenness centrality \cite{Ulusoy.2018} &   &   &   & 1 &  1 &    &  1 &    &             &    &    &  1 &    &   4 \\
                                          composite resilience metric \cite{S.S.Ottenburger.2019} &   &   &   & 1 &  1 &    &  1 &    &             &  1 &    &  1 &    &   4 \\
                                                      overall system resilience \cite{Sweya.2020} &   &   & 1 & 1 &  1 &    &  1 &    &           1 &  1 &    &  1 &    &   4 \\
                                                             resilience indicator \cite{Rak.2020} &   &   &   & 1 &  1 &    &    &    &           1 &    &    &    &    &   4 \\
                                                 weighted average path length \cite{DiNardo.2018} &   &   &   & 1 &  1 &    &  1 &    &             &    &    &  1 &    &   4 \\
                                                    water provision resilience \cite{Milman.2008} & 1 & 1 &   & 1 &    &  1 &    &    &           1 &  1 &  1 &  1 &  1 &   5 \\
\hline
\end{longtblr}

\endgroup

\bibliographystyle{elsarticle-num}
\bibliography{bibliography.bib}

\begin{thebibliography}{10}
\expandafter\ifx\csname url\endcsname\relax
  \def\url#1{\texttt{#1}}\fi
\expandafter\ifx\csname urlprefix\endcsname\relax\def\urlprefix{URL }\fi
\expandafter\ifx\csname href\endcsname\relax
  \def\href#1#2{#2} \def\path#1{#1}\fi

\bibitem{UnitedNationsSustainableDevelopment.1142022}
{United Nations Sustainable Development},
  \href{https://www.un.org/sustainabledevelopment/water-and-sanitation/}{Water
  and sanitation - united nations sustainable development} (11/4/2022).
\newline\urlprefix\url{https://www.un.org/sustainabledevelopment/water-and-sanitation/}

\bibitem{UNDESA.July2022}
{UN DESA}, \href{https://unstats.un.org/sdgs/report/2022/}{The sustainable
  development goals report 2022} (2022).
\newline\urlprefix\url{https://unstats.un.org/sdgs/report/2022/}

\bibitem{Koks.2022}
E.~E. Koks, K.~C.~H. {van Ginkel}, M.~J.~E. {van Marle}, A.~Lemnitzer, Brief
  communication: Critical infrastructure impacts of the 2021 mid-july western
  european flood event, Natural Hazards and Earth System Sciences 22~(12)
  (2022) 3831--3838.
\newblock \href {https://doi.org/10.5194/nhess-22-3831-2022}
  {\path{doi:10.5194/nhess-22-3831-2022}}.

\bibitem{Shumilova.2023}
O.~Shumilova, K.~Tockner, A.~Sukhodolov, V.~Khilchevskyi, L.~de~Meester,
  S.~Stepanenko, G.~Trokhymenko, J.~A. Hern{\'a}ndez-Ag{\"u}ero, P.~Gleick,
  Impact of the russia--ukraine armed conflict on water resources and water
  infrastructure, Nature Sustainability (2023).
\newblock \href {https://doi.org/10.1038/s41893-023-01068-x}
  {\path{doi:10.1038/s41893-023-01068-x}}.

\bibitem{MiddleEastEye.3282023}
{Middle East Eye},
  \href{https://www.middleeasteye.net/news/turkey-earthquake-lack-clean-water-toilets-survivors-risk-disease}{Turkey
  earthquake: Lack of clean water and toilets puts survivors at risk of
  disease} (3/28/2023).
\newline\urlprefix\url{https://www.middleeasteye.net/news/turkey-earthquake-lack-clean-water-toilets-survivors-risk-disease}

\bibitem{Ulusoy.2018}
A.-J. Ulusoy, I.~Stoianov, A.~Chazerain, {Hydraulically informed graph
  theoretic measure of link criticality for the resilience analysis of water
  distribution networks}, {Applied network science} 3~(1) (2018) 31.
\newblock \href {https://doi.org/10.1007/s41109-018-0079-y}
  {\path{doi:10.1007/s41109-018-0079-y}}.

\bibitem{Fekete.2020}
A.~Fekete, T.~Hartmann, R.~J{\"u}pner,
  \href{https://wires.onlinelibrary.wiley.com/doi/full/10.1002/wat2.1397}{Resilience:
  On--going wave or subsiding trend in flood risk research and practice?},
  WIREs Water 7~(1) (2020) e1397.
\newblock \href {https://doi.org/10.1002/wat2.1397}
  {\path{doi:10.1002/wat2.1397}}.
\newline\urlprefix\url{https://wires.onlinelibrary.wiley.com/doi/full/10.1002/wat2.1397}

\bibitem{Liu.2020}
W.~Liu, Z.~Song, {Review of studies on the resilience of urban critical
  infrastructure networks}, {Reliability Engineering {\&} System Safety} 193
  (2020) 106617.
\newblock \href {https://doi.org/10.1016/j.ress.2019.106617}
  {\path{doi:10.1016/j.ress.2019.106617}}.

\bibitem{Shuang.2019}
Q.~Shuang, H.~J. Liu, E.~Porse, {Review of the Quantitative Resilience Methods
  in Water Distribution Networks}, {Water} 11~(6) (2019) 1189.
\newblock \href {https://doi.org/10.3390/w11061189}
  {\path{doi:10.3390/w11061189}}.

\bibitem{Shin.2018}
S.~Shin, S.~Lee, D.~Judi, M.~Parvania, E.~Goharian, T.~McPherson, S.~Burian, {A
  Systematic Review of Quantitative Resilience Measures for Water
  Infrastructure Systems}, {Water} 10~(2) (2018) 164.
\newblock \href {https://doi.org/10.3390/w10020164}
  {\path{doi:10.3390/w10020164}}.

\bibitem{Gunawan.2017}
I.~Gunawan, F.~Schultmann, S.~A. Zarghami, The four rs performance indicators
  of water distribution networks, International Journal of Quality {\&}
  Reliability Management 34~(5) (2017) 720--732.
\newblock \href {https://doi.org/10.1108/IJQRM-11-2016-0203}
  {\path{doi:10.1108/IJQRM-11-2016-0203}}.

\bibitem{Gay.2013}
L.~F. Gay, S.~K. Sinha, {Resilience of civil infrastructure systems: literature
  review for improved asset management}, {International Journal of Critical
  Infrastructures} 9~(4) (2013) 330.
\newblock \href {https://doi.org/10.1504/IJCIS.2013.058172}
  {\path{doi:10.1504/IJCIS.2013.058172}}.

\bibitem{Mohebbi.2020}
S.~Mohebbi, Q.~Zhang, E.~{Christian Wells}, T.~Zhao, H.~Nguyen, M.~Li,
  N.~Abdel-Mottaleb, S.~Uddin, Q.~Lu, M.~J. Wakhungu, Z.~Wu, Y.~Zhang,
  A.~Tuladhar, X.~Ou, {Cyber-physical-social interdependencies and
  organizational resilience: A review of water, transportation, and cyber
  infrastructure systems and processes}, {Sustainable Cities and Society} 62
  (2020) 102327.
\newblock \href {https://doi.org/10.1016/j.scs.2020.102327}
  {\path{doi:10.1016/j.scs.2020.102327}}.

\bibitem{Mohebbi.2021}
S.~Mohebbi, K.~Barnett, B.~Aslani, {Decentralized resource allocation for
  interdependent infrastructures resilience: a cooperative game approach},
  {International Transactions in Operational Research} 28~(6) (2021)
  3394--3415.
\newblock \href {https://doi.org/10.1111/itor.12978}
  {\path{doi:10.1111/itor.12978}}.

\bibitem{Holling.1973}
C.~S. Holling, {Resilience and Stability of Ecological Systems}, {Annual Review
  of Ecology and Systematics} 4~(1) (1973) 1--23.
\newblock \href {https://doi.org/10.1146/annurev.es.04.110173.000245}
  {\path{doi:10.1146/annurev.es.04.110173.000245}}.

\bibitem{Holling.1996}
C.~S. Holling, {Engineering Resilience versus Ecological Resilience}, in: P.~C.
  Schulze (Ed.), {Engineering within Ecological Constraints}, {National
  Academies Press}, Washington, D.C., USA, 1996, pp. 31--43.

\bibitem{Chandler.2016}
D.~Chandler, J.~Coaffee,
  \href{https://www.taylorfrancis.com/books/9781315765006}{{The Routledge
  Handbook of International Resilience}}, {Routledge handbooks}, {Taylor and
  Francis}, Florence, 2016.
\newline\urlprefix\url{https://www.taylorfrancis.com/books/9781315765006}

\bibitem{Elsner.2018}
I.~Elsner, A.~Huck, M.~Marathe, {Resilience}, in: J.~I. Engels (Ed.), {Key
  Concepts for Critical Infrastructure Research}, {Springer Fachmedien
  Wiesbaden}, Wiesbaden, 2018, pp. 31--38.

\bibitem{Canizares.2021}
J.~C. Ca{\~n}izares, S.~M. Copeland, N.~Doorn,
  \href{https://www.mdpi.com/2071-1050/13/15/8538}{{Making Sense of
  Resilience}}, {Sustainability} 13~(15) (2021) 8538.
\newblock \href {https://doi.org/10.3390/su13158538}
  {\path{doi:10.3390/su13158538}}.
\newline\urlprefix\url{https://www.mdpi.com/2071-1050/13/15/8538}

\bibitem{Woods.2017}
D.~Woods, E.~Hollnagel,
  \href{https://ebookcentral.proquest.com/lib/kxp/detail.action?docID=5118484}{Resilience
  Engineering: Concepts and Precepts}, 1st Edition, {CRC Press} and Safari,
  Erscheinungsort nicht ermittelbar and Boston, MA, 2017.
\newline\urlprefix\url{https://ebookcentral.proquest.com/lib/kxp/detail.action?docID=5118484}

\bibitem{Hollnagel.2011}
E.~Hollnagel (Ed.), {Resilience Engineering in Practice: A Guidebook}, {Ashgate
  studies in resilience engineering}, Ashgate, Farnham, 2011.

\bibitem{Hollnagel.2016}
E.~Hollnagel,
  \href{https://erikhollnagel.com/ideas/resilience\%20assessment\%20grid.html}{Resilience
  assessment grid} (30.07.2016).
\newline\urlprefix\url{https://erikhollnagel.com/ideas/resilience\%20assessment\%20grid.html}

\bibitem{Pelz.2021b}
P.~F. Pelz, P.~Groche, M.~E. Pfetsch, M.~Schaeffner, {Mastering Uncertainty in
  Mechanical Engineering}, {Springer International Publishing}, Cham, 2021.
\newblock \href {https://doi.org/10.1007/978-3-030-78354-9}
  {\path{doi:10.1007/978-3-030-78354-9}}.

\bibitem{Hosseini.2016}
S.~Hosseini, K.~Barker, J.~E. Ramirez-Marquez, A review of definitions and
  measures of system resilience, Reliability Engineering {\&} System Safety 145
  (2016) 47--61.
\newblock \href {https://doi.org/10.1016/j.ress.2015.08.006}
  {\path{doi:10.1016/j.ress.2015.08.006}}.

\bibitem{Forster.2015}
B.~F{\"o}rster, M.~Bauch (Eds.), Wasserinfrastrukturen und Macht von der Antike
  bis zur Gegenwart, Vol.~63 of Historische Zeitschrift // Beihefte (Neue
  Folge), {DE GRUYTER}, 2015.
\newblock \href {https://doi.org/10.1515/9783486781052}
  {\path{doi:10.1515/9783486781052}}.

\bibitem{Moss.2020}
T.~Moss,
  \href{https://ebookcentral.proquest.com/lib/kxp/detail.action?docID=6340832}{Remaking
  Berlin: A history of the city through infrastructure, 1920-2020},
  Infrastructures series, {The MIT Press}, Cambridge, Massachusetts and London,
  England, 2020.
\newline\urlprefix\url{https://ebookcentral.proquest.com/lib/kxp/detail.action?docID=6340832}

\bibitem{EuropeanCommitteeforStandardization.April2022}
{European Committee for Standardization}, Water supply - requirements for
  systems and components outside buildings (April 2022).

\bibitem{DINNormenausschussWasserwesen.February2021}
{Deutsches Institut f{\"u}r Normung e. V.}, Guidelines for the management of
  assets of water supply and wastewater systems -- part 1: Drinking water
  distribution networks (iso 24516-1:2016) (February 2021).

\bibitem{BundesamtfurBevolkerungsschutzundKatastrophenhilfe.22.02.2022}
{Bundesamt f{\"u}r Bev{\"o}lkerungsschutz und Katastrophenhilfe},
  \href{https://www.bbk.bund.de/SharedDocs/Downloads/DE/Mediathek/Publikationen/KRITIS/rahmenkonzept-trinkwassernotversorgung.pdf?{\_}{\_}blob=publicationFile\&v=1}{Rahmenkonzept
  der trinkwassernotversorgung: Neukonzeption zur anpasssung an ver{\"a}nderte
  rahmenbedingunen in anlehnung an die konzeption zivile verteidigung (2016)}
  (2022).
\newline\urlprefix\url{https://www.bbk.bund.de/SharedDocs/Downloads/DE/Mediathek/Publikationen/KRITIS/rahmenkonzept-trinkwassernotversorgung.pdf?{\_}{\_}blob=publicationFile\&v=1}

\bibitem{WesternCapeGovernment.20.04.2023}
{Western Cape Government},
  \href{https://www.westerncape.gov.za/general-publication/cape-town-water-rationing}{Cape
  town water rationing} (2018).
\newline\urlprefix\url{https://www.westerncape.gov.za/general-publication/cape-town-water-rationing}

\bibitem{McCann.2018}
M.~McCann, C.~Knudsen, \href{https://spherestandards.org/}{The Sphere handbook:
  Humanitarian charter and minimum standards in humanitarian response}, fourth
  edition Edition, {Sphere Association} and {Practical Action Publishing}, Genf
  and Rugby, 2018.
\newline\urlprefix\url{https://spherestandards.org/}

\bibitem{Sweya.2021}
L.~N. Sweya, S.~Wilkinson, {Tool development to measure the resilience of water
  supply systems in Tanzania: Economic dimension}, {J{\`a}mb{\'a} - Journal of
  Disaster Risk Studies} 13~(1) (2021).
\newblock \href {https://doi.org/10.4102/jamba.v13i1.860}
  {\path{doi:10.4102/jamba.v13i1.860}}.

\bibitem{Mentges.2023}
A.~Mentges, L.~Halekotte, M.~Schneider, T.~Demmer, D.~Lichte,
  \href{https://arxiv.org/pdf/2302.04524}{{A resilience glossary shaped by
  context: Reviewing resilience-related terms for critical infrastructures}}
  (2023).
\newline\urlprefix\url{https://arxiv.org/pdf/2302.04524}

\bibitem{Cassottana.2021}
B.~Cassottana, N.~Y. Aydin, L.~C. Tang, Quantitative assessment of system
  response during disruptions: An application to water distribution systems,
  Journal of Water Resources Planning and Management 147~(3) (2021).
\newblock \href {https://doi.org/10.1061/(ASCE)WR.1943-5452.0001334}
  {\path{doi:10.1061/(ASCE)WR.1943-5452.0001334}}.

\bibitem{Huizar.2018}
L.~H. Huizar, K.~E. Lansey, R.~G. Arnold, {Sustainability, robustness, and
  resilience metrics for water and other infrastructure systems}, {Sustainable
  and Resilient Infrastructure} 3~(1) (2018) 16--35.
\newblock \href {https://doi.org/10.1080/23789689.2017.1345252}
  {\path{doi:10.1080/23789689.2017.1345252}}.

\bibitem{HernandezHernandez.2021}
E.~{Hernandez Hernandez}, L.~Ormsbee, {Segment-Based Assessment of Consequences
  of Failure on Water Distribution Systems}, {Journal of Water Resources
  Planning and Management} 147~(4) (2021) 04021009.
\newblock \href {https://doi.org/10.1061/(ASCE)WR.1943-5452.0001340}
  {\path{doi:10.1061/(ASCE)WR.1943-5452.0001340}}.

\bibitem{Dubaniowski.2021}
M.~I. Dubaniowski, H.~R. Heinimann, {Framework for modeling interdependencies
  between households, businesses, and infrastructure system, and their response
  to disruptions---application}, {Reliability Engineering {\&} System Safety}
  212 (2021) 107590.
\newblock \href {https://doi.org/10.1016/j.ress.2021.107590}
  {\path{doi:10.1016/j.ress.2021.107590}}.

\bibitem{Milman.2008}
A.~Milman, A.~Short, {Incorporating resilience into sustainability indicators:
  An example for the urban water sector}, {Global Environmental Change} 18~(4)
  (2008) 758--767.
\newblock \href {https://doi.org/10.1016/j.gloenvcha.2008.08.002}
  {\path{doi:10.1016/j.gloenvcha.2008.08.002}}.

\bibitem{Izadi.2020}
A.~Izadi, F.~Yazdandoost, R.~Ranjbar, {Asset-Based Assessment of Resiliency in
  Water Distribution Networks}, {Water Resources Management} 34~(4) (2020)
  1407--1422.
\newblock \href {https://doi.org/10.1007/s11269-020-02508-5}
  {\path{doi:10.1007/s11269-020-02508-5}}.

\bibitem{Cubillo.2019}
F.~Cubillo, {\'A}.~Mart{\'i}nez-Codina, {A metric approach to measure
  resilience in water supply systems}, {Journal of Applied Water Engineering
  and Research} 7~(1) (2019) 67--78.
\newblock \href {https://doi.org/10.1080/23249676.2017.1355758}
  {\path{doi:10.1080/23249676.2017.1355758}}.

\bibitem{Zhang.2020}
Q.~Zhang, F.~Zheng, Q.~Chen, Z.~Kapelan, K.~Diao, K.~Zhang, Y.~Huang,
  {Improving the Resilience of Postdisaster Water Distribution Systems Using
  Dynamic Optimization Framework}, {Journal of Water Resources Planning and
  Management} 146~(2) (2020) 04019075.
\newblock \href {https://doi.org/10.1061/(ASCE)WR.1943-5452.0001164}
  {\path{doi:10.1061/(ASCE)WR.1943-5452.0001164}}.

\bibitem{Krueger.2019}
E.~H. Krueger, D.~Borchardt, J.~W. Jawitz, H.~Klammler, S.~Yang, J.~Zischg,
  P.~S.~C. Rao, {Resilience Dynamics of Urban Water Supply Security and
  Potential of Tipping Points}, {Earth's Future} 7~(10) (2019) 1167--1191.
\newblock \href {https://doi.org/10.1029/2019EF001306}
  {\path{doi:10.1029/2019EF001306}}.

\bibitem{Cimellaro.2016b}
G.~P. Cimellaro, A.~Tinebra, C.~Renschler, M.~Fragiadakis, {New Resilience
  Index for Urban Water Distribution Networks}, {Journal of Structural
  Engineering} 142~(8) (2016).
\newblock \href {https://doi.org/10.1061/(ASCE)ST.1943-541X.0001433}
  {\path{doi:10.1061/(ASCE)ST.1943-541X.0001433}}.

\bibitem{Kong.2019}
J.~Kong, S.~P. Simonovic, C.~Zhang, {Resilience Assessment of Interdependent
  Infrastructure Systems: A Case Study Based on Different Response Strategies},
  {Sustainability} 11~(23) (2019) 6552.
\newblock \href {https://doi.org/10.3390/su11236552}
  {\path{doi:10.3390/su11236552}}.

\bibitem{Farahmandfar.2018}
Z.~Farahmandfar, K.~R. Piratla, {Comparative Evaluation of Topological and
  Flow-Based Seismic Resilience Metrics for Rehabilitation of Water Pipeline
  Systems}, {Journal of Pipeline Systems Engineering and Practice} 9~(1) (2018)
  04017027.
\newblock \href {https://doi.org/10.1061/(ASCE)PS.1949-1204.0000293}
  {\path{doi:10.1061/(ASCE)PS.1949-1204.0000293}}.

\bibitem{Ren.2020}
K.~Ren, S.~Huang, Q.~Huang, H.~Wang, G.~Leng, W.~Fang, P.~Li, {Assessing the
  reliability, resilience and vulnerability of water supply system under
  multiple uncertain sources}, {Journal of Cleaner Production} 252 (2020)
  119806.
\newblock \href {https://doi.org/10.1016/j.jclepro.2019.119806}
  {\path{doi:10.1016/j.jclepro.2019.119806}}.

\bibitem{Zhuang.2013}
B.~Zhuang, K.~Lansey, D.~Kang, {Resilience/Availability Analysis of Municipal
  Water Distribution System Incorporating Adaptive Pump Operation}, {Journal of
  Hydraulic Engineering} 139~(5) (2013) 527--537.
\newblock \href {https://doi.org/10.1061/(ASCE)HY.1943-7900.0000676}
  {\path{doi:10.1061/(ASCE)HY.1943-7900.0000676}}.

\bibitem{Hashimoto.1982}
T.~Hashimoto, J.~R. Stedinger, D.~P. Loucks, {Reliability, resiliency, and
  vulnerability criteria for water resource system performance evaluation},
  {Water Resources Research} 18~(1) (1982) 14--20.
\newblock \href {https://doi.org/10.1029/WR018i001p00014}
  {\path{doi:10.1029/WR018i001p00014}}.

\bibitem{Zhao.2017}
S.~Zhao, X.~Liu, Y.~Zhuo, {Hybrid Hidden Markov Models for resilience metrics
  in a dynamic infrastructure system}, {Reliability Engineering {\&} System
  Safety} 164 (2017) 84--97.
\newblock \href {https://doi.org/10.1016/j.ress.2017.02.009}
  {\path{doi:10.1016/j.ress.2017.02.009}}.

\bibitem{ElBaroudy.2004}
I.~El-Baroudy, S.~P. Simonovic, {Fuzzy criteria for the evaluation of water
  resource systems performance}, {Water Resources Research} 40~(10) (2004).
\newblock \href {https://doi.org/10.1029/2003WR002828}
  {\path{doi:10.1029/2003WR002828}}.

\bibitem{Nasrazadani.2020}
H.~Nasrazadani, M.~Mahsuli, {Probabilistic Framework for Evaluating Community
  Resilience: Integration of Risk Models and Agent-Based Simulation}, {Journal
  of Structural Engineering} 146~(11) (2020) 04020250.
\newblock \href {https://doi.org/10.1061/(ASCE)ST.1943-541X.0002810}
  {\path{doi:10.1061/(ASCE)ST.1943-541X.0002810}}.

\bibitem{Liu.2020b}
W.~Liu, Z.~Song, M.~Ouyang, J.~Li, {Recovery-based seismic resilience
  enhancement strategies of water distribution networks}, {Reliability
  Engineering {\&} System Safety} 203 (2020) 107088.
\newblock \href {https://doi.org/10.1016/j.ress.2020.107088}
  {\path{doi:10.1016/j.ress.2020.107088}}.

\bibitem{Yazdani.2012}
A.~Yazdani, P.~Jeffrey, {Water distribution system vulnerability analysis using
  weighted and directed network models}, {Water Resources Research} 48~(6)
  (2012).
\newblock \href {https://doi.org/10.1029/2012WR011897}
  {\path{doi:10.1029/2012WR011897}}.

\bibitem{DiNardo.2018}
A.~{Di Nardo}, M.~{Di Natale}, C.~Giudicianni, R.~Greco, G.~F. Santonastaso,
  {Complex network and fractal theory for the assessment of water distribution
  network resilience to pipe failures}, {Water Supply} 18~(3) (2018) 767--777.
\newblock \href {https://doi.org/10.2166/ws.2017.124}
  {\path{doi:10.2166/ws.2017.124}}.

\bibitem{Herrera.2016}
M.~Herrera, E.~Abraham, I.~Stoianov, A graph-theoretic framework for assessing
  the resilience of sectorised water distribution networks, Water Resources
  Management 30~(5) (2016) 1685--1699.
\newblock \href {https://doi.org/10.1007/s11269-016-1245-6}
  {\path{doi:10.1007/s11269-016-1245-6}}.

\bibitem{Dwivedi.2014}
A.~Dwivedi, {Designing for resilience}, in: I.~V. Ternovskiy, P.~Chin (Eds.),
  {Cyber Sensing 2014}, {SPIE Proceedings}, SPIE, 2014, p. 90970C.
\newblock \href {https://doi.org/10.1117/12.2054389}
  {\path{doi:10.1117/12.2054389}}.

\bibitem{Balaei.2018}
B.~Balaei, S.~Wilkinson, R.~Potangaroa, N.~Hassani, M.~Alavi-Shoshtari,
  {Developing a Framework for Measuring Water Supply Resilience}, {Natural
  Hazards Review} 19~(4) (2018) 04018013.
\newblock \href {https://doi.org/10.1061/(ASCE)NH.1527-6996.0000292}
  {\path{doi:10.1061/(ASCE)NH.1527-6996.0000292}}.

\bibitem{Rak.2020}
J.~R. Rak, J.~{\.Z}ywiec, {Selected Aspects of the Water Supply System Safety},
  in: Z.~Blikharskyy, P.~Koszelnik, P.~Mesaros (Eds.), {Proceedings of CEE
  2019}, Vol.~47 of {Lecture Notes in Civil Engineering}, {Springer
  International Publishing}, Cham, 2020, pp. 369--376.
\newblock \href {https://doi.org/10.1007/978-3-030-27011-7{\textunderscore }47}
  {\path{doi:10.1007/978-3-030-27011-7{\textunderscore }47}}.

\bibitem{Sweya.2020}
L.~N. Sweya, S.~Wilkinson, J.~Mayunga, A.~Joseph, G.~Lugomela, J.~Victor,
  {Development of a Tool to Measure Resilience against Floods for Water Supply
  Systems in Tanzania}, {Journal of Management in Engineering} 36~(4) (2020)
  05020007.
\newblock \href {https://doi.org/10.1061/(ASCE)ME.1943-5479.0000783}
  {\path{doi:10.1061/(ASCE)ME.1943-5479.0000783}}.

\bibitem{S.S.Ottenburger.2019}
S.~S. Ottenburger, S.~Bai, W.~Raskob, {MCDA-based Genetic Algorithms for
  Developing Disaster Resilient Designs of Critical Supply Networks}, in:
  Y.~Hadjadj-Aoul (Ed.), {The 6th International Conference on Information and
  Communication Technologies for Disaster Management}, IEEE, Piscataway, NJ,
  2019, pp. 1--4.
\newblock \href {https://doi.org/10.1109/ICT-DM47966.2019.9032982}
  {\path{doi:10.1109/ICT-DM47966.2019.9032982}}.

\bibitem{Assad.2019}
A.~Assad, O.~Moselhi, T.~Zayed, {A New Metric for Assessing Resilience of Water
  Distribution Networks}, {Water} 11~(8) (2019) 1701.
\newblock \href {https://doi.org/10.3390/w11081701}
  {\path{doi:10.3390/w11081701}}.

\bibitem{Wright.2015}
R.~Wright, M.~Herrera, P.~Parpas, I.~Stoianov, {Hydraulic Resilience Index for
  the Critical Link Analysis of Multi-feed Water Distribution Networks},
  {Procedia Engineering} 119 (2015) 1249--1258.
\newblock \href {https://doi.org/10.1016/j.proeng.2015.08.987}
  {\path{doi:10.1016/j.proeng.2015.08.987}}.

\bibitem{TANYIMBOH.1993}
T.~T. TANYIMBOH, A.~B. TEMPLEMAN, {Optimum Design of Flexible Water
  Distribution Networks}, {Civil Engineering Systems} 10~(3) (1993) 243--258.
\newblock \href {https://doi.org/10.1080/02630259308970126}
  {\path{doi:10.1080/02630259308970126}}.

\bibitem{DiNardo.2015}
A.~{Di Nardo}, M.~{Di Natale}, G.~F. Santonastaso, V.~G. Tzatchkov, V.~H.
  Alcocer-Yamanaka, {Performance indices for water network partitioning and
  sectorization}, {Water Supply} 15~(3) (2015) 499--509.
\newblock \href {https://doi.org/10.2166/ws.2014.132}
  {\path{doi:10.2166/ws.2014.132}}.

\bibitem{Jeong.2020}
G.~Jeong, D.~Kang, {Hydraulic Uniformity Index for Water Distribution
  Networks}, {Journal of Water Resources Planning and Management} 146~(2)
  (2020) 04019078.
\newblock \href {https://doi.org/10.1061/(ASCE)WR.1943-5452.0001158}
  {\path{doi:10.1061/(ASCE)WR.1943-5452.0001158}}.

\bibitem{Gonzales.2017}
P.~Gonzales, N.~K. Ajami, {An integrative regional resilience framework for the
  changing urban water paradigm}, {Sustainable Cities and Society} 30 (2017)
  128--138.
\newblock \href {https://doi.org/10.1016/j.scs.2017.01.012}
  {\path{doi:10.1016/j.scs.2017.01.012}}.

\bibitem{Sirsant.2020}
S.~Sirsant, M.~J. Reddy, {Assessing the Performance of Surrogate Measures for
  Water Distribution Network Reliability}, {Journal of Water Resources Planning
  and Management} 146~(7) (2020) 04020048.
\newblock \href {https://doi.org/10.1061/(ASCE)WR.1943-5452.0001244}
  {\path{doi:10.1061/(ASCE)WR.1943-5452.0001244}}.

\bibitem{He.2019}
X.~He, Y.~Yuan, {A Framework of Identifying Critical Water Distribution
  Pipelines from Recovery Resilience}, {Water Resources Management} 33~(11)
  (2019) 3691--3706.
\newblock \href {https://doi.org/10.1007/s11269-019-02328-2}
  {\path{doi:10.1007/s11269-019-02328-2}}.

\bibitem{EzioTodini.2000}
{Ezio Todini}, {Looped water distribution networks design using a resilience
  index based heuristic approach}, {Urban Water Journal} 2~(2) (2000) 115--122.
\newblock \href {https://doi.org/10.1016/S1462-0758(00)00049-2}
  {\path{doi:10.1016/S1462-0758(00)00049-2}}.

\bibitem{Altherr.2018}
L.~C. Altherr, N.~Br{\"o}tz, I.~Dietrich, T.~Gally, F.~Ge{\ss}ner,
  H.~Kloberdanz, P.~Leise, P.~F. Pelz, P.~D. Schlemmer, A.~Schmitt, {Resilience
  in Mechanical Engineering - A Concept for Controlling Uncertainty during
  Design, Production and Usage Phase of Load-Carrying Structures}, {Applied
  Mechanics and Materials} 885 (2018) 187--198.
\newblock \href {https://doi.org/10.4028/www.scientific.net/AMM.885.187}
  {\path{doi:10.4028/www.scientific.net/AMM.885.187}}.

\bibitem{Creaco.2016b}
E.~Creaco, M.~Franchini, E.~Todini, {Generalized Resilience and Failure Indices
  for Use with Pressure-Driven Modeling and Leakage}, {Journal of Water
  Resources Planning and Management} 142~(8) (2016) 04016019.
\newblock \href {https://doi.org/10.1061/(ASCE)WR.1943-5452.0000656}
  {\path{doi:10.1061/(ASCE)WR.1943-5452.0000656}}.

\bibitem{Liu.2014}
H.~Liu, D.~Savi{\'c}, Z.~Kapelan, M.~Zhao, Y.~Yuan, H.~Zhao, {A
  diameter-sensitive flow entropy method for reliability consideration in water
  distribution system design}, {Water Resources Research} 50~(7) (2014)
  5597--5610.
\newblock \href {https://doi.org/10.1002/2013WR014882}
  {\path{doi:10.1002/2013WR014882}}.

\bibitem{Farahmandfar.2017}
Z.~Farahmandfar, K.~R. Piratla, R.~D. Andrus, {Resilience Evaluation of Water
  Supply Networks against Seismic Hazards}, {Journal of Pipeline Systems
  Engineering and Practice} 8~(1) (2017) 04016014.
\newblock \href {https://doi.org/10.1061/(ASCE)PS.1949-1204.0000251}
  {\path{doi:10.1061/(ASCE)PS.1949-1204.0000251}}.

\bibitem{CubidesCastro.2021}
E.~Cubides-Castro, C.~L{\'o}pez-Aburto, P.~Iglesias-Rey,
  F.~Mart{\'i}nez-Solano, D.~Mora-Meli{\'a}, M.~Iglesias-Castell{\'o},
  {Methodology for Determining the Maximum Potentially Recoverable Energy in
  Water Distribution Networks}, {Water} 13~(4) (2021) 464.
\newblock \href {https://doi.org/10.3390/w13040464}
  {\path{doi:10.3390/w13040464}}.

\bibitem{BinMahmoud.2018}
A.~A. {Bin Mahmoud}, K.~R. Piratla, {Comparative evaluation of resilience
  metrics for water distribution systems using a pressure driven demand-based
  reliability approach}, {Journal of Water Supply: Research and
  Technology-Aqua} 67~(6) (2018) 517--530.
\newblock \href {https://doi.org/10.2166/aqua.2018.010}
  {\path{doi:10.2166/aqua.2018.010}}.

\bibitem{Amarasinghe.2016}
P.~Amarasinghe, Liu, P.~Egodawatta, P.~Barnes, J.~McGree, A.~Goonetilleke,
  {Quantitative assessment of resilience of a water supply system under
  rainfall reduction due to climate change}, {Journal of Hydrology} 540 (2016)
  1043--1052.
\newblock \href {https://doi.org/10.1016/j.jhydrol.2016.07.021}
  {\path{doi:10.1016/j.jhydrol.2016.07.021}}.

\bibitem{Eppstein.1998}
D.~Eppstein, Finding the k shortest paths, SIAM Journal on Computing 28~(2)
  (1998) 652--673.
\newblock \href {https://doi.org/10.1137/S0097539795290477}
  {\path{doi:10.1137/S0097539795290477}}.

\bibitem{Lorenz.2020}
I.-S. Lorenz, P.~Pelz, Optimal resilience enhancement of water distribution
  systems, Water 12~(9) (2020) 2602.
\newblock \href {https://doi.org/10.3390/w12092602}
  {\path{doi:10.3390/w12092602}}.

\bibitem{LuisAngelGarciaEscudero.2003}
{Luis Angel Garc{\'i}a-Escudero}, {Alfonso Gordaliza}, {Carlos Matr{\'a}n},
  Trimming tools in exploratory data analysis, Journal of Computational and
  Graphical Statistics 12~(2) (2003) 434--449.
\newblock \href {https://doi.org/10.1198/1061860031806}
  {\path{doi:10.1198/1061860031806}}.

\bibitem{Ostfeld.2016}
A.~Ostfeld, \href{https://uknowledge.uky.edu/wdst\_models/2}{04 calibration
  networks}, Battle of the Water Network Models 2 (2016).
\newline\urlprefix\url{https://uknowledge.uky.edu/wdst\_models/2}

\bibitem{Folke.2006}
C.~Folke,
  \href{https://www.sciencedirect.com/science/article/pii/S0959378006000379}{Resilience:
  The emergence of a perspective for social--ecological systems analyses},
  Global Environmental Change 16~(3) (2006) 253--267.
\newblock \href {https://doi.org/10.1016/j.gloenvcha.2006.04.002}
  {\path{doi:10.1016/j.gloenvcha.2006.04.002}}.
\newline\urlprefix\url{https://www.sciencedirect.com/science/article/pii/S0959378006000379}

\bibitem{Bogardi.2019}
J.~J. Bogardi, A.~Fekete,
  \href{https://cos.bibl.th-koeln.de/frontdoor/index/index/docId/862}{From
  intriguing concept(s) towards an overused buzzword: is it time for a requiem
  for resilience?}, in: K.~Fehn, A.~Fekete, C.~Hetk{\"a}mper, A.~Lechleuthner,
  C.~Norf, O.~A. Mudimu, U.~Schremmer (Eds.), Resilience and Vulnerability:
  Conceptual revolution(s) or only revolving around words? A collection of
  essays, working papers and think pieces from the period 2008-2018, Vol.
  03/2019 of Integrative Risk and Security Research, {Bibliothek der
  Technischen Hochschule K{\"o}ln}, K{\"o}ln, 2019, pp. 70--87.
\newline\urlprefix\url{https://cos.bibl.th-koeln.de/frontdoor/index/index/docId/862}

\bibitem{Cai.2020}
Y.~Cai, Renaissance of resilience: A buzzword or a new ideal?, Management and
  Organization Review 16~(5) (2020) 976--980.
\newblock \href {https://doi.org/10.1017/mor.2020.46}
  {\path{doi:10.1017/mor.2020.46}}.

\end{thebibliography}
\end{document}